\documentclass[11pt,a4paper]{article}

\usepackage[utf8]{inputenc}
\usepackage[T1]{fontenc}
\usepackage{lmodern} % Classic, elegant serif typeface
\usepackage{graphicx}
\usepackage{epstopdf}
\usepackage{amsmath}
\usepackage{amssymb}
\usepackage{amsfonts}
\usepackage{cite}    % Standard bracket style:, [2]
\usepackage{booktabs} 
\usepackage[numbers,sort&compress]{natbib}
\usepackage[top=1.2in, bottom=1.2in, left=.8in, right=.8in]{geometry}

\usepackage{indentfirst}    % Indent the first paragraph of every section
\usepackage{titlesec}
\titleformat{\section}[block]{\normalfont\filcenter\scshape\large}{\thesection.}{0.5em}{}
\titlespacing*{\section}{0pt}{24pt}{12pt}

\titleformat{\subsection}[block]{\normalfont\itshape}{\thesubsection.}{0.5em}{}
\titlespacing*{\subsection}{0pt}{18pt}{6pt}

\newcommand{\beq}{\begin{equation}}
\newcommand{\eeq}{\end{equation}}
\newcommand{\pfr}[2]{\ensuremath{\frac{\partial #1}{\partial #2}}}

\newcommand{\Pec}{\textit{Pe}}
\newcommand{\Rey}{\textit{Re}}
\newcommand{\Sch}{\textit{Sc}}

\newcommand{\LL}{\mathcal{L}}

\newcommand{\mb}[1]{\mathbf{#1}}

\newcommand{\gb}[1]{\boldsymbol{#1}}
\newcommand{\mc}[1]{\mathcal{#1}}
\makeatletter
\def\@maketitle{%
  \newpage
  \null
  \vskip 2em%
  \begin{center}%
  {\LARGE\bfseries \@title \par}% Title: Bold, elegant serif
  \vskip 1.5em%
  {\large \scshape \@author \par}% Author: Small caps
  \vskip 1em%
  {\small \@date}% Date / Communicated by info
  \end{center}%
  \par
  \vskip 1.5em}
\makeatother

\renewenvironment{abstract}{%
  \begin{center}{\scshape\small Abstract}\end{center}%
  \small\itshape
  \begin{quotation}%
}{%
  \end{quotation}%
  \vskip 2em
}

\usepackage{fancyhdr}
\title{Taylor dispersion and large-deviation structure in stagnation-point flows without homogenisation}
\author{Prabakaran Rajamanickam \\ \small \textit{Department of Mathematics \& Statistics, University of Strathclyde, Glasgow G1 1XQ, UK}}
\begin{document}

\maketitle

\begin{abstract}
We study the long-time dispersion of a passive scalar in linearly accelerating flows whose strain rate $v(\eta)$ varies transversely: the axially strained flow in
a porous tube with wall injection, and radially strained stagnation-point flows over a
wall (Homann and von K\'arm\'an flows). Exponential stretching outpaces cross-sectional
diffusion, so the scalar never homogenises across the section and classical
cross-sectional averaging fails. Instead, in the logarithm of the stretched coordinate
the problem becomes a Taylor dispersion problem with a Boltzmann cross-section, with mean log-stretching
rate $U=\Lambda'(0)$ and log-space diffusivity $D=\tfrac{1}{2}\Lambda''(0)$: Taylor dispersion emerges
without homogenisation, and only because the strain rate varies transversely. A Mellin-transform analysis, based on the principal eigenvalue $\Lambda(q)$ of the
transverse advection-diffusion operator augmented by $qv$, gives the large-time
structure in terms of the log-speed $x=\ln|\boldsymbol\xi|/t$, where $\boldsymbol\xi\in \mathbb R^d$ denotes the stretching directions ($d=1$ axial and $d=2$ radial). Essentially all the scalar mass
lies in a log-normal bulk at $x=U$, flanked by forward and rear large-deviation tails
with mass density $e^{-I(x)t}$, where $I$ is the rate function (i.e., the Legendre transform of $\Lambda(q)$). Behind
the rear tail, below $x_*=\Lambda'(-d)<U$, a plateau independent of
$\gb\xi$ decays at the rate $-\Lambda(-d)$. It carries the highest
concentrations but only a fraction $e^{-I(x_*)t}$ of the mass. A thin transition zone
 joins the plateau to the rear tail. For uniformly
strained flows, as in the Burgers vortex, all regions collapse onto a single Gaussian.
\end{abstract}

\section{Introduction}\label{sec:intro}

Taylor dispersion~\cite{taylor1953dispersion} is the classic example of how a
transversely varying velocity, together with cross-sectional diffusion, turns a localised
solute cloud into an effectively diffusing one. The theory rests on a separation of
time scales: once the cloud is spread much longer than the transverse diffusion time permits,
the concentration is nearly uniform across the section, and only the section average
needs to be tracked. The idea has since been extended to many unidirectional, periodic, porous
and certain generalised laminar flows~\cite{watson1983diffusion,erdogan1967effects,
prigozhin1982solute,frankel1989foundations,brenner1980dispersion,majda1999simplified,
rajamanickam2023effective,rajamanickam2025shear,rajamanickam2025taylor}.

Stagnation-point-type flows break this separation. Whether the flow is confined in a
porous tube with uniform wall injection~\cite{yuan1956laminar,taylor1956fluid,
culick1966rotational}, generated by an accelerating wall~\cite{brady1981steady}, or
formed over a solid surface, the velocity grows linearly with distance from the
stagnation region. A scalar cloud is then stretched exponentially in time, at a rate
that varies across the section, and diffusion across the section cannot keep up. The
cloud does not homogenise, and an effective axial diffusivity based on cross-sectional
averages is not defined. Standard Taylor dispersion
analysis, which requires coefficients that are periodic in, or independent of the longitudinal position, therefore is not applicable here since the velocity grows linearly with $\boldsymbol\xi$. Translation invariance is restored
only in $\ln|\boldsymbol\xi|$. Such flows arise near counterflow flames and mixing
layers, in injected tubes such as the Taylor--Culick model of propellant chambers, and
in boundary layers where a scalar is released near a stagnation point.

Exact solutions are available when the strain rate is uniform. A point source in a linear
flow keeps a Gaussian profile whose variance, depending on time, obeys linear ordinary differential
equations. This was obtained for steady pure strain by
\cite{townsend1951diffusion} and \cite{batchelor1979mass}, for general
two-dimensional linear flows, including simple shear and rotation, by
\cite{foister1980diffusion}, and for time-dependent strain by
\cite{krishnan1992diffusion}. In the steady planar case the variance along the
stretching axis grows like $e^{2t}$ while the transverse variance saturates.
These solutions rely on the strain being the same at every point, so that diffusion
factorises along the principal axes and, in the logarithm of the stretched coordinate,
the cloud is simply translated at the strain rate without spreading. This is the exact
analogue of a uniform flow in classical Taylor dispersion, which only translates the
solute; dispersion requires a velocity that varies across the section. Once the
strain rate depends on the transverse coordinate and transverse diffusion exchanges
material between streamlines, the factorisation is lost, and none of these solutions
applies.

Two earlier treatments addressed the setting for non-uniform strain rates. \cite{chatwin1974dispersion}
considered the planar flow near a forward stagnation point. For large times he assumed
that the transverse profile is the same at every streamwise position, as in the exact
solution for uniform strain, and replaced the strain rate by a transverse average. This
gives a Gaussian cloud stretched at a single effective rate. It is an approximation
rather than a derivation: it locates the bulk of the scalar, but not the spreading of
the cloud or the structure behind it. The present author~\cite{rajamanickam2020dispersion} considered
axisymmetric flows (counterflow, Homann and von K\'arm\'an), took the stretching to
dominate at large times, and found that near the axis the concentration approaches a
profile that is uniform in the stretched coordinate and decays at a rate given by an
eigenvalue of a transverse operator, which differs from Chatwin's profile. This is
recovered here as the plateau region. That solution was, however, written as a single
Gaussian-type profile for the whole cloud, which is not uniformly valid and does not
conserve mass, and for the von K\'arm\'an flow the eigenvalue reported there is not the
asymptotic decay rate.

The present paper gives a complete large-time description. Our central step is to use
$\ln|\boldsymbol\xi|$ as the longitudinal coordinate, where
$\boldsymbol\xi\in\mathbb R^d$ is the stretched coordinate ($d=1$ for axially strained
flows and $d=2$ for radially strained flows). Far from the release point, longitudinal
molecular diffusion is negligible and the transport equation becomes exactly a Taylor
dispersion problem in this logarithmic variable, with a transverse velocity equal to the
local strain rate $v(\eta)$, where $\eta$ is the transverse coordinate. The
cross-section relaxes to a Boltzmann profile $m_\infty(\eta)$, and the logarithmic
position of a solute molecule behaves as a biased random walk with mean log-stretching
rate $U$ and Taylor diffusivity $D$, both given by quadratures over $m_\infty$. As in the
classical problem, $D>0$ only if $v$ varies across the section; for uniform $v$ one has
$D=0$ and the Gaussian solution of \cite{townsend1951diffusion} is recovered. Taylor
dispersion therefore emerges here without any homogenisation, but only in the logarithm
of the stretched coordinate, and only because the strain varies transversely.

\begin{figure}[h!]
\centering
\includegraphics[scale=0.85]{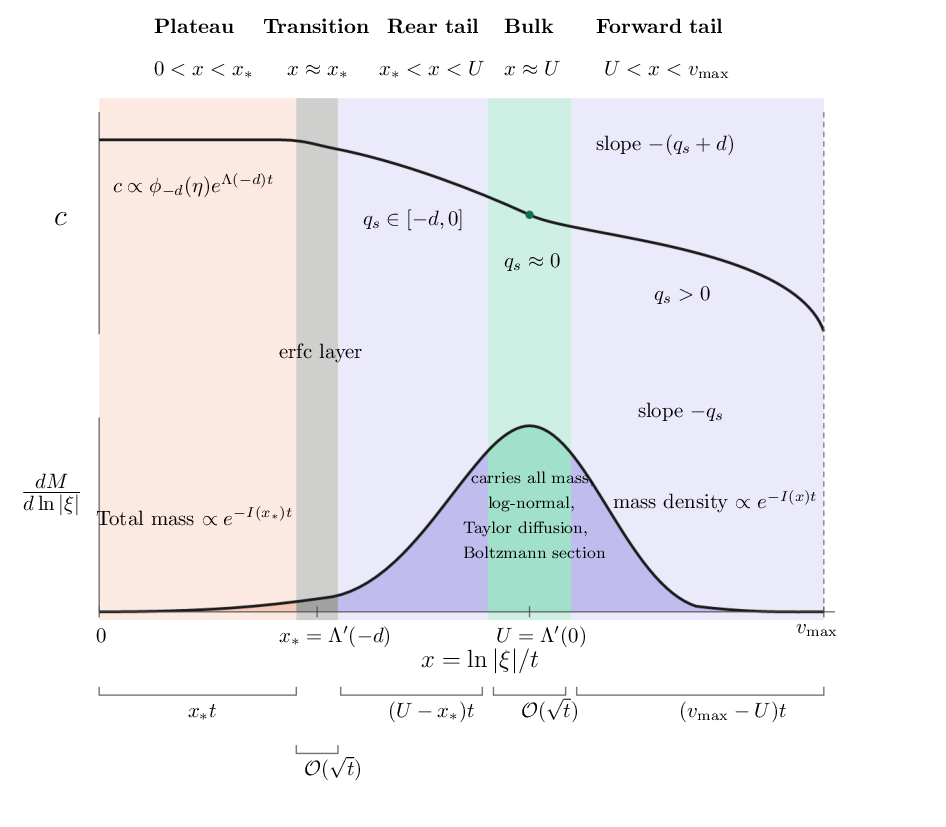}\vspace{-0.6cm}
\caption{Schematic (not to scale) of the large-time structure of the scalar cloud as a
function of the logarithmic speed $x=\ln|\boldsymbol\xi|/t$. Top: concentration $c$ at
fixed $\eta$ on a logarithmic scale. Bottom: scalar mass per unit $\ln|\boldsymbol\xi|$,
integrated over $\eta$. The shaded zones are, from left to right, the plateau, the
transition zone, the rear tail, the bulk and the forward tail. Here $q_s(x)$ is defined
by $\Lambda'(q_s)=x$, and $I(x)=q_sx-\Lambda(q_s)$ is the rate function. The slope of
$\ln c$ against $\ln|\boldsymbol\xi|$ is $-(q_s+d)$ and the slope of the logarithm of the
mass per unit $\ln|\boldsymbol\xi|$ is $-q_s$. The widths of the zones in
$\ln|\boldsymbol\xi|$ are indicated below the axis.}
\label{fig:sch}
\end{figure}

This Taylor-dispersion description is, however, only the centre of the picture. A
Mellin transform in $|\boldsymbol\xi|$ leads to the principal eigenvalue $\Lambda(q)$ of
the transverse advection--diffusion operator $\mathcal L$ plus $qv$, which is the
cumulant generating function of the accumulated log-stretch, with $U=\Lambda'(0)$ and
$D=\Lambda''(0)/2$. A saddle-point analysis then shows that the scalar cloud has
a richer structure than a single log-normal bulk, and that the answer to the question
\emph{where is the scalar?} depends on whether one asks about concentration or about
scalar mass. The structure is summarised in figure~\ref{fig:sch}, in terms of the speed
$x=\ln|\boldsymbol\xi|/t$ in logarithmic space. Let $q_s(x)$ be defined by
$\Lambda'(q_s)=x$ and $I(x)=q_sx-\Lambda(q_s)$, the large-deviation rate function.
Starting from where the mass is, and moving towards the stagnation point, the cloud
consists of:
\begin{itemize}
\item a \emph{bulk} at $\ln|\boldsymbol\xi|=Ut$ with width of order $\sqrt t$, which
carries essentially all the mass in a log-normal profile of variance $2Dt$ in
$\ln|\boldsymbol\xi|$ with the Boltzmann cross-section $m_\infty(\eta)$;
\item \emph{forward and rear large-deviation tails} on either side of the bulk, where
the mass per unit $\ln|\boldsymbol\xi|$ decays like $e^{-I(x)t}$. The cross-section
changes continuously along the tails ($q_s\in[-d,0]$ on the rear side and $q_s>0$ on
the forward side), and the full function $\Lambda(q)$, not just $U$ and $D$, is needed.
The forward tail terminates at the maximum strain rate $v_{\max}$. The rear tail
extends back to $x_*=\Lambda'(-d)<U$. Both tails have a width of order $t$ in
$\ln|\boldsymbol\xi|$;
\item a \emph{transition zone} at $x=x_*$, a complementary-error-function layer of width
of order $\sqrt t$, which joins the rear tail to the plateau;
\item a \emph{plateau} behind it ($0<x<x_*$), where the concentration is independent of
$\boldsymbol\xi$ and decays as $e^{\Lambda(-d)t}$ with cross-section $\phi_{-d}(\eta)$,
the eigenfunction of $\mathcal L-dv$. The plateau has the highest concentration in
the cloud, exceeding the peak bulk concentration by the exponentially large factor
$e^{[\Lambda(-d)+dU]t}$, yet its total mass is only of order $e^{-I(x_*)t}$.
\end{itemize}
The plateau therefore describes what a probe near the stagnation point records, while
the bulk describes where most of the scalar is, and the two are exponentially far apart,
separated by $(U-x_*)t$ in $\ln|\boldsymbol\xi|$. The two kinds of region differ in size
as well: the bulk and transition zone have width of order $\sqrt t$, while the plateau
and the tails grow linearly in time. The plateau recovers the axis result of
\cite{rajamanickam2020dispersion}, but the other regions, and the way mass is
distributed among them, were not previously known. For uniform strain, $D=0$ and
$U=x_*=v$, so the tails and the transition zone collapse and the structure merges into
the known Gaussian solution. Here, it is worth noting that there have been recent developments~\cite{haynes2014dispersion,haynes2014dispersion2,tzella2016dispersion,kahlen2017large} in the large-deviation regime of Taylor dispersion, which better explain the tails of concentration fields in parallel shear and periodic cellular channels. However, those frameworks fundamentally assume an underlying separation of timescales where cross-sectional homogenisation eventually holds. By contrast, the multi-region structural architecture derived here demonstrates that accelerating stagnation-point flows entirely break this classical homogenization, forcing an entirely distinct, full-field Eulerian evolution.

We also treat two points that affect the tube. First, with injection through the wall,
a no-flux condition on the concentration alone would make the wall a scalar source;
the correct condition is that the total flux vanishes, which is a Robin condition.
Second, the plateau decay rate is a property of the whole cross-section and is not the
maximum strain rate. For the von K\'arm\'an flow, $\Lambda(-d)$ is replaced by the edge
of the continuous spectrum, and the previously quoted decay rate is not the
asymptotic one.

The remainder of the paper is organised as follows. Section~\ref{sec:formulation}
formulates the problem for both geometries in a unified form, and
Section~\ref{sec:axially} derives the Boltzmann cross-section, whereas Section~\ref{sec:taylor} derives the Taylor dispersion in logarithmic coordinates.
Section~\ref{sec:triple} derives the full structure described above: the bulk, the
large-deviation tails, the transition zone and the plateau. Section~\ref{sec:structural} provides some structural consequences of the large-time solution. Section~\ref{sec:Lambda} examines the spectral property of the transverse operator $\mc L+qv$. The numerical method and its
comparison with the asymptotic results are described in Section~\ref{sec:numerics}.

\section{Problem formulation and unified framework}
\label{sec:formulation}

\subsection{Axially strained flow in a tube}
Consider the scalar-transport problem in a slender porous tube of circular cross-section with radius $a$ and length $2L$ with $a/L\ll 1$. Fluid is uniformly injected into the tube with a speed $V>0$. Both ends of the tube are assumed open, although the analysis also applies if only one end is open. The characteristic radial velocity scales as $v_r^*\sim V$, while the axial velocity scales as $v_z^*\sim V L/a$~\cite{linan2004flow}. In terms of the non-dimensional variables
\begin{align}
    (r,z)=\frac{1}{a}(r^*,z^*),  \qquad (v_r,v_z)=\frac{1}{V}(v_r^*,v_z^*), \qquad p= \frac{p^*}{\rho V^2},
\end{align}
the self-similar solution is given by
\begin{equation}
    v_r = -\frac{2}{r}f(\eta), \qquad v_z= 4z f'(\eta), \qquad p = -\frac{1}{2}(\gamma z)^2 + p_0(\eta)
\end{equation}
where $\eta=r^2$ and a prime denotes differentiation with respect to $\eta$. The function $f'(\eta)$ represents the \textit{local axial strain rate} at different radial locations. Substituting the above ansatz into the axial momentum equation yields the nonlinear eigenvalue problem~\cite{yuan1956laminar}
\begin{equation}
    \Rey^{-1}(\eta f'''+f'') + ff'' - f'^2 = -\tfrac{1}{16}\gamma^2 \label{feq}
\end{equation}
subject to the boundary conditions
\begin{equation}
    f(0)=0, \qquad \lim_{\eta\to 0}\sqrt{\eta}f''(\eta)=0, \qquad f(1)=\tfrac{1}{2}, \qquad f'(1)=0. \label{fbc}
\end{equation}
Here, $\Rey=Va/\nu$ is the Reynolds number and the coefficient $\gamma$--representing the axial pressure gradient--must be determined as part of the solution. The radial pressure distribution $p_0(\eta)$ is obtained by integrating the radial momentum equation,
\begin{equation}
    p_0' = \frac{2f^2}{\eta^2} - \frac{4ff'}{\eta} - \frac{4f''}{\Rey}.
\end{equation}
As $\Rey\to 0$, the flow approaches a Poiseuille-type flow, whereas as $\Rey\to \infty$, it tends to the Taylor--Culick flow~\cite{balachandar2001generation}, i.e.,
\begin{align}
    f = \eta - \tfrac{1}{2}\eta^2 + \mc O(\Rey), \qquad \gamma\sqrt{\Rey} = 4 + \mc O(\Rey) \quad \text{as} \quad \Rey\to 0,  \label{Poi}\\
    f =  \tfrac{1}{2}\sin \frac{\pi\eta}{2} + \mc O(\Rey^{-1}), \qquad \gamma = \pi + \mc O(\Rey^{-1}) \quad \text{as} \quad \Rey\to \infty. \label{TC}
\end{align}
\cite{durlofsky1984spatial} demonstrated that the problem~\eqref{feq}-\eqref{fbc} admits a unique solution for all $\Rey>0$. The description of the flow field for the analogous problem driven by a linearly stretching wall can be formulated in a similar manner~\cite{brady1981steady}; for brevity, we shall just focus on the Taylor--Culick-type flow.

We now introduce a compact distribution of passive scalar with concentration $c=c_0(r,z)$ in the flow field described above. The equation governing $c$ is given by
\begin{equation}
    \pfr{c}{t} + \Pec \left(v_r \pfr{c}{r}+v_z\pfr{c}{z}\right) = \frac{1}{r}\pfr{}{r}\left(r\pfr{c}{r}\right) + \pfr{^2c}{z^2}
\end{equation}
where $t=t^*D_c/a^2$ is the dimensionless time, $\Pec=Va/D_c=\Rey\Sch$ is the Péclet number and $\Sch=\nu/D_c$ is the Schmidt number. Writing the governing equation, in conservation form using $\nabla\cdot\mb v$,  and introducing $\xi=z$, we have
\begin{equation}
    \pfr{c}{t}=4\pfr{}{\eta}\left(\eta\pfr{c}{\eta}+\Pec fc\right) - 4\Pec f'\pfr{(\xi c)}{\xi}+\pfr{^2c}{\xi^2}. \label{eq:tube}
\end{equation}
The total radial scalar flux is $J=-4(\eta c_\eta+\Pec fc)$. Since $v_r(1)=-1\ne0$, setting a simple Neumann condition, $c_\eta=0$, at the wall, implies an advective flux $-2\Pec c$ leaking across the boundary. Thus, a strict no-flux condition with a scalar-free injected fluid is that total scalar flux vanishes at the wall, establishing the boundary condition
\begin{equation}
     \pfr{c}{\eta}+\frac{1}{2}\Pec\,  c=0 \qquad \text{at} \qquad \eta=1, \label{eq:robin}
\end{equation}
while maintaining the axis regularity condition $\sqrt{\eta}c_\eta \to 0$ as $\eta \to 0$. The global scalar conservation requires
\begin{equation}
    M_{\rm tot}\equiv\int_{-\infty}^{\infty}\!\!\int_0^1 c(\eta,\xi,t)\,\mathrm d\eta\,\mathrm d\xi=1 . \label{conservation}
\end{equation}
For the tube the physical amount of scalar is $\pi M_{\rm tot}$ (the area element is $\pi d\eta$); since the problem is linear, $M_{\rm tot}=1$ is only normalisation.

\subsection{Radially strained flow over a solid surface}

The dispersion problem in radially strained flows has been formulated in~\cite{rajamanickam2020dispersion}. Following this, the fluid velocity field scales as $v_r^*=kr^*f'(\eta)$ and $v_z^*=-2\sqrt{k\nu}f(\eta)$, where $k>0$ is the strain/rotation rate and $\eta=z^*/\sqrt{\nu/k}$ is the non-dimensional distance from the solid surface. In the Homann flow, $f(\eta)$ is governed by the problem
\begin{align}
    f'''+2ff''-f'^2+1=0, \qquad f(0)=f'(0)=f'(\infty)-1=0;
\end{align}
in the von K\'arm\'an flow, it satisfies the problem
\begin{align}
    f'''+2ff''-f'^2+g^2=0, &\qquad f(0)=f'(0)=f'(\infty)=0,\\
    g'' + 2fg'-2f'g=0 , &\qquad g(0)-1=g(\infty)=0,
\end{align}
where $v_\theta^*=kr^*g(\eta)$. Setting the stretched radial coordinate $\xi=r^*/\sqrt{\nu/k}$ and dimensionless time $t=t^*D_ck/\nu$, the conservative transport equation, assuming axisymmetry, expands on $\eta>0$ as
\begin{equation}
    \pfr{c}{t}=\pfr{}{\eta}\left(\pfr{c}{\eta}+2\Sch f c\right) -\frac{\Sch}{\xi}\pfr{(\xi^2f'c)}{\xi}+\frac{1}{\xi}\pfr{}{\xi}\left(\xi\pfr{c}{\xi}\right). \label{eq:open}
\end{equation}
Crucially, because $f(0)=0$, the normal fluid velocity vanishes exactly at the solid interface ($v_z=0$ at $\eta=0$). Consequently, there is no convective flux crossing the solid surface, and the standard Neumann condition $c_\eta=0$ remains perfectly valid and physically consistent.

\subsection{Axisymmetric counterflow with uniform strain rate: e.g. Burgers vortex}\label{sec:burgers} 

As a benchmark case, we include the axially strained flow with a uniform strain rate such as the Burgers vortex. The meridional components of the Burgers vortex are $v_r^*=-kr^*$ and
$v_z^*=2kz^*$, with $k>0$. Since only the strain rate $k$ and the diffusivity
$D_c$ enter the scalar problem, we set $r=r^*/\sqrt{D_c/k}$, $\xi=z^*/\sqrt{D_c/k}$,
$t=kt^*$ and $\eta=r^2$, so that the transport equation becomes
\begin{equation}
    \pfr{c}{t}=4\pfr{}{\eta}\left[\eta\left(\pfr{c}{\eta}+\frac{1}{2}c\right)\right]
    -2\pfr{(\xi c)}{\xi}+\pfr{^2c}{\xi^2}.
    \label{eq:burgers}
\end{equation}

\subsection{Unified coordinate form}
Let $\gb\xi\in\mathbb R^d$ denote a generalized vector tracking the stretched coordinate directions ($d=1$ representing the axial position $z$ inside the tube; $d=2$ tracking the radial plane in the unconfined flow). Both configurations collapse perfectly onto the single unified equation
\begin{equation}
    c_t=\mc L[c]-v(\eta)\,\nabla_{\xi}\cdot(\gb\xi c)+\delta\,\Delta_\xi c,\qquad \LL[c]=\partial_\eta\left[\alpha(\eta)\,(c_\eta+\beta(\eta)c)\right], \label{eq:unified}
\end{equation}
where $\mc L[c]$ represents the transverse advection-diffusion operator acting across the cross-section, $\nabla_{\xi}\cdot(\gb\xi c)$ represents the advective stretching term along the expanding coordinate lines and $\Delta_\xi$ is the Laplacian operator acting purely in the stretching directions. The advective stretching term $-v(\eta)\,\nabla_{\xi}\cdot(\gb\xi c)= - vdc - v\gb\xi\cdot\nabla_\xi c$ captures the local cloud dilution due to volume expansion and spatial advection via a velocity field that accelerates linearly with distance. The normalisation is given by $M_{\rm tot}=\int_{\Omega_{\eta}}\int_{\mathbb R^d}c d^d\xi d\eta$. The geometry-specific transport parameters of the two problems are itemized in Table~\ref{tab:data}.

\begin{table}
\centering\small
\begin{tabular}{lccc}
\toprule
 & axially strained tube flow & radially strained surface flow & Burgers \\
\midrule
domain, $\Omega_\eta$ & $[0,1]$ & $[0,\infty)$ & $[0,\infty)$ \\
$\alpha(\eta)$, $\beta(\eta)$ & $4\eta$, $\ \Pec f/\eta$ & $1$, $\ 2\Sch f$ & $4\eta$, $\tfrac12$ \\
strain rate $v(\eta)$ & $4\Pec f'$ & $\Sch f'$ & 2\\
stretched dimension $d$, $\delta$ & $1$, $1$ & $2$, $1$ & $1$, $1$ \\
\bottomrule
\end{tabular}
\caption{Ingredients of the unified transport relation \eqref{eq:unified}.}
\label{tab:data}
\end{table}

From here onwards, we use the terminology \textit{longitudinal} for the stretching directions $(\gb\xi)$ and \textit{transverse} to the remaining directions $(\eta)$. Throughout, the strain rate is non-negative, $v(\eta)\ge0$, so that the flow
stretches the scalar everywhere in the cross-section. This holds for the  two problems exhibiting transverse variations in the strain rate  considered
here (Table~\ref{tab:data}).

\section{Longitudinally integrated field and its eigenfunctions}
\label{sec:axially}

We begin by examining the behaviour of the unified transport relation~\eqref{eq:unified} integrated over the entire stretching domain $\mathbb{R}^d$. Let $m(\eta, t) = \int_{\Omega_\xi} c \, d^d\xi$ define the longitudinally integrated scalar mass profile. Integrating~\eqref{eq:unified} causes both the advective divergence and the longitudinal molecular diffusion terms to vanish under zero-flux or decaying boundaries at infinity, leaving a closed one-dimensional relation for the transverse distribution,
\begin{equation}
    m_t = \mc L[m] = \partial_\eta \!\big[ \alpha(\eta) \left( m_\eta + \beta(\eta) m \right) \big].
\end{equation}
At large times $(t\gg 1)$, the solution to this problem can be written as
\begin{equation}
    m(\eta,t)=m_\infty + \sum_{n=1}^\infty a_n e^{-\lambda_n t}m_n(\eta)
\end{equation}
where $(m_n,\lambda_n)$ obeys the eigenvalue problem
\begin{equation}
    \mc L[m_n]=-\lambda_n m_n \quad \text{satisfying the condition } J=0 \text{ on transverse boundaries.}
\end{equation}
Over long timescales ($t \gg 1$), the higher-order transverse modes decay exponentially, and the system relaxes completely to the unique stationary transverse profile $m_\infty(\eta)$ corresponding to the zero eigenvalue ($\LL m_\infty = 0$). Imposing a zero net-flux condition across the transverse boundaries reveals that the stationary profile satisfies the first-order balance $m_\eta + \beta(\eta)m = 0$, yielding a standard Boltzmann distribution,
\begin{equation}\label{boltzmann}
    m_\infty(\eta) = \frac{\exp\left( -\int_0^\eta \beta(s) \, d s \right)}{\int_{\Omega_\eta} \exp\left( -\int_0^\eta \beta(s) \, d s \right) d\eta},
\end{equation}
so that $\int_{\Omega_\eta} m_\infty  d\eta = 1$. In the absence of transverse convection $(\beta=0)$ as in the classical Poiseuille flow, $m_\infty=1$ becomes a constant.

As itemized in Table~\ref{tab:data}, for the axially strained flow in the tube, we have $m_\infty \propto \exp\left(-\Pec \int_0^\eta [f(s)/s]\,d s\right)$. This implies, as $\Pec\to \infty$, we can replace $f(s)\approx f'(0)s$  with the implication that $m_\infty(\eta)\propto \exp(-\Pec f'(0)\eta)$, indicating that the scalar gets trapped in a thin boundary layer of width $\eta \sim 1/\Pec$ right along the axis. The approximation $f(s)\approx f'(0)s$ is essentially nothing but the axially strained stagnation-point flow (such as the Burgers' vortex) since $v_r\approx -2f'(0)r$ and $v_z\approx 4zf'(0)$; see Appendix~\ref{app:largepe} for details.  On the other hand, the dispersion in radial strained flows, already resides in the intrinsic boundary-layer system, it does not require any asymptotic limit to break free of the finite confinement (tube radius for the first problem).

The relaxation of $m$ towards $m_\infty$ is governed by the non-zero eigenvalues of the non-self-adjoint transverse operator $\mathcal{L}$. The formal adjoint operator $\mathcal{L}^\dagger$ determined via the $L^2$ inner product identity $\langle \mathcal{L}[m], g \rangle = \langle m, \mathcal{L}^\dagger[g] \rangle$ leads to the eigenvalue problem
\begin{equation}\label{eq:relax}
    \mc L^\dagger[g_n]=-\lambda_n g_n \quad \text{satisfying the condition } g'=0 \text{ on transverse boundaries,}
\end{equation}
where $\mc L^\dagger[g]=\alpha(\eta)g''+[\alpha'(\eta)-\alpha(\eta)\beta(\eta)]g'$.
 Crucially, the non-zero eigenvalues $\lambda_n$ govern the internal transverse
relaxation towards the Boltzmann state $m_\infty$.

\section{Logarithmic coordinate transformation and Taylor dispersion}\label{sec:taylor}

At large times, the longitudinal extent of the cloud is exponentially large. At this scale, the longitudinal molecular diffusion term ($\delta \Delta_\xi c$) is (exponentially) subdominant and can be ignored in the first instance. To see this directly, write the unified equation in Fourier space (for the tube problem, imposing even symmetry in $z$) and in Hankel space of order zero (for the surface problem; $\hat c=\int_0^\infty\xi J_0(k\xi)c\,d\xi$). In both cases,\footnote{With $\hat c=\int_0^\infty\xi J_0(k\xi)c\,d\xi$, integration by parts gives $\int\xi J_0(k\xi)\,\xi c_\xi\,d\xi=-2\hat c-k\hat c_k$, so $-\xi f'c_\xi$ transforms to $f'(2\hat c+k\hat c_k)$; combined with $2fc_\eta=2(fc)_\eta-2f'c$ (and the corresponding conservation form) one obtains the stated equation with $v=\Sch f'$, $\delta=1$.} we obtain
\begin{equation}
     \hat c_t =\mc L[\hat c]  + v(\eta)k\hat c_k - \delta k^2\hat c.
\end{equation}
To resolve the exponentially large distances $(k\to 0)$, we introduce $Y=\ln k$, so that
\begin{equation} \label{eq:chat}
    \hat c_t =\mc L[\hat c]  + v(\eta)\hat c_Y - \delta e^{2Y}\hat c.
\end{equation}
Taking $Y\to -\infty$ keeps all terms intact except the  longitudinal diffusion term. From here onwards, we neglect this diffusion term and will justify this in Section~\ref{sec:justify} and show that its correction is exponentially negligible.

 To track the spreading cloud at these large distances ($|\gb\xi| \gg 1$), we introduce the logarithmic variables
\begin{equation}
    X = \ln|\gb\xi|, \qquad N = |\gb\xi|^d c.
\end{equation}
Since the differential scalar mass $dM=c dV= N S_ddX d\eta$ ($S_d$ is the surface area of the unit sphere in $\mathbb R^d$), because $dV=S_d |\gb\xi|^{d-1}d|\gb\xi|d\eta=S_d|\gb\xi|^d dXd\eta$ (the volume element in $(\gb\xi,\eta)$ space; for the tube the physical one is $\pi$ times this), $N(X, \eta, t)$ represents the local scalar mass per unit longitudinal logarithm and unit transverse coordinate,  such that $S_d\int_{-\infty}^{\infty} N \, d X = m(\eta, t)$. For ($|\gb\xi| \gg 1$), ignoring the longitudinal diffusion, the governing equation simplifies to
\begin{equation}
    N_t = \LL[N] - v(\eta) N_X. \label{eq:log_transport}
\end{equation}
This equation is a classical Taylor dispersion problem for large Peclet numbers without molecular diffusion along $X$; the function $v(\eta)$ corresponds to a steady, transversally varying  longitudinal velocity in the logarithmic variable, and $\LL$ corresponds to the net transverse flux. Thus, Taylor-dispersion effects emerge for the scalar-mass quantity $N(X,\eta,t)$ rather than the concentration $c(X,\eta,t)$, as a direct consequence of the exponential stretching of the fluid-volume elements.

To solve the transport relation~\eqref{eq:log_transport}, we introduce the moments
\begin{equation}
    \mc M_q(\eta,t)=\int_{\Omega_\xi}|\gb\xi|^q c\, d^d\xi = S_d\int_{-\infty}^\infty e^{qX} N\, dX,
\end{equation}
with $\mc M_0=m(\eta,t)$. From the viewpoint of $N$, $\mc M_q$ is a moment generating function. The Taylor-dispersion regime itself needs only its behaviour near $q=0$, i.e.\ the first two cumulants. To describe the cloud beyond that regime, however, we need solutions for arbitrary $q$, and in particular $q=-d$, as shown below. We therefore retain the full generating function, which evolves as
\begin{equation}
    \partial_t \mc M_q = (\LL + q v)\,\mc M_q, \label{eq:moments}
\end{equation}
closed for each $q$. At large times, the behaviour is controlled by the principal eigenvalue $\Lambda(q)$ of the operator $\LL + q v$,
\begin{equation}
    (\LL + q v)\phi_q = \Lambda(q)\phi_q, \label{eq:Lambda}
\end{equation}
with $\phi_q(\eta)$ is the principal eigenfunction, subject to the normalization condition $\int\phi_q\,d\eta=1$. The function $\Lambda(q)$ is the cumulant generating function of the log-stretching accumulated along a tracer path. Expanding this problem about $q=0$ recovers the Taylor-dispersion quantities. Specifically, we write
\begin{align}
    \Lambda(q)&=Uq+D q^2+\dots, \qquad \text{with} \qquad \Lambda(0)=0, \label{eq:eigexp}\\
    \phi_q&= m_\infty + q \psi_1 +q^2\psi_2+\cdots \qquad \text{with} \qquad \phi_{0}=m_\infty,
\end{align}
so that $(\LL+qv)(m_\infty+q\psi_1+\dots)=(qU+q^2D+\dots)(m_\infty+q\psi_1+\dots)$. At $\mc O(q)$, we have
\begin{equation}
    \LL[\psi_1]=(U-v)m_\infty.
\end{equation}
Integrating this over the domain using the transverse boundary condition leads to the expression for the mean log-stretching rate, 
\begin{equation} \label{eq:U}
    U = \Lambda'(0)=\int_{\Omega_\eta} v(\eta) m_\infty(\eta)d\eta.
\end{equation}
The function $\psi_1(\eta)$ is solved (see Appendix~\ref{app:taylor}) to be 
\begin{equation}
    \psi_1=m_\infty h_1,\qquad h_1'=-\frac{W(\eta)}{\alpha(\eta)\,m_\infty(\eta)},\qquad
W(\eta)=\int_0^\eta[v(s)-U]\,m_\infty(s)\,\mathrm ds ,
\end{equation}
This function satisfies the required normalisation condition $\int_{\Omega_\eta} \psi_1d\eta=0$ and the no-flux condition $\psi_1'+\beta\psi_1=0$ on $\partial{\Omega_\eta}$, arising at this order. At $\mc O(q^2)$, the eigenvalue equation simplifies to
\begin{equation}
    \LL[\psi_2]  = (U-v)\psi_1 +  D m_\infty
\end{equation}
Integrating this over the domain yields
\begin{equation}
    D = \int_{\Omega_\eta} \frac{W'(\eta)}{m_\infty(\eta)}\psi_1(\eta)d\eta.
\end{equation}
Upon integration of parts, this reduces to the following expression for the Taylor diffusivity of $\ln|\gb\xi|$,
\begin{equation} \label{eq:D}
    D = \tfrac{1}{2}\Lambda''(0)= \int_{\Omega_\eta}\frac{W(\eta)^2}{\alpha(\eta)\,m_\infty(\eta)}\,d\eta,
\end{equation}
which ensures the positive definiteness of $D$. Equations~\eqref{eq:U} and~\eqref{eq:D} have the same form as in classical Taylor dispersion, with $v$ playing the role of the axial velocity. They differ in two respects. First, the weight $m_\infty$ is not uniform: the transverse flow ($\beta\neq0$) makes it Boltzmann-like, so $U$ is neither the area-averaged nor the maximum strain rate. Second, $D$ has no molecular contribution, since longitudinal diffusion is negligible here, and it describes the spreading of $\ln|\boldsymbol\xi|$ rather than of position.

\section{The asymptotic structure at large times}
\label{sec:triple}

\subsection{Mellin transform and the saddle point}
The large-time structure follows from the Mellin transform of the concentration with respect to the longitudinal coordinate magnitude $|\gb\xi|$,
\begin{equation}
    \tilde c(\eta,q,t)=\int_0^\infty c(\eta,|\gb\xi|,t)\,|\gb\xi|^{q+d-1}\,d|\gb\xi|,\qquad q\in\mathbb C,
    \label{eq:mellin_def}
\end{equation}
which is related to the moments~\eqref{eq:moments} by
\begin{equation}
    \mc M_q=S_d\,\tilde c(\eta,q,t)
    \label{eq:M_c}
\end{equation}
The part $|\gb\xi|^{d-1}$ in the Mellin transform comes from the area element. Neglecting longitudinal diffusion (justified in Section~\ref{sec:justify}), $\tilde c$ obeys the same equation as $\mc M_q$,
\begin{equation}
    \partial_t\tilde c=\left(\LL+qv\right)\tilde c .
    \label{eq:mellin_eq}
\end{equation}
At large times its solution is dominated by the principal eigenvalue $\Lambda(q)$ and the corresponding right eigenfunction $\phi_q(\eta)$ of $\LL+qv$, so that
\begin{equation}
    \tilde c(\eta,q,t)=a(q)\,e^{\Lambda(q)t}\,\phi_{q}(\eta),
    \label{eq:mellin_sol}
\end{equation}
where $a(q)$ is the integration constant and the only piece of the formula that carries memory of the early diffusive stage. Since $c$ decays rapidly as $|\gb\xi|\to\infty$, the integral~\eqref{eq:mellin_def} converges at large $|\gb\xi|$ for all $q$, whereas at small $|\gb\xi|$ it converges only for $\mathrm{Re}\{q+d\}>0$. Hence $\tilde c$ is analytic in $\mathrm{Re}\{q+d\}>0$. Because $c$ is smooth and non-zero at $\gb\xi=\mathbf 0$, we can write
\begin{equation} \label{eq:cexpansion}
    c=c_0(\eta,t) + c_2(\eta,t)|\gb\xi|^2+c_4(\eta,t)|\gb\xi|^4+\cdots \quad \text{as} \quad |\gb\xi|\to 0.
\end{equation}
Therefore the integral near the origin behaves like $c_0\int_0\!|\gb\xi|^{q+d-1}d|\gb\xi|=c_0/(q+d) + \cdots$. Thus $\tilde c$ has simple poles at $q+d=0,-2,-4,\dots$ with residues $c_0,c_2,\dots$. The leading-order structure is dominated by the simple pole at $q=-d$ with residue $c_0$. Considering only this pole, we isolate the singular part by writing
\begin{equation}
    a(q)=\frac{K}{q+d}+\breve a(q),
\end{equation}
where $\breve a(q)$ is regular near $q=-d$ and $K$ is the residue of $a$ at $q=-d$. The inverse Mellin transform is taken along a vertical line $\mathrm{Re}\{q+d\}=\sigma>0$, to the right of the pole,
\begin{equation}
    c(\eta,|\gb\xi|,t)=\frac1{2\pi i}\int_{\sigma-i\infty}^{\sigma+i\infty}a(q)\,e^{\Lambda(q)t}\,\phi_{q}(\eta)\,|\gb\xi|^{-(q+d)}\,dq.
    \label{eq:mellin}
\end{equation}

To analyse $t\gg1$ at exponentially large distances we set $|\gb\xi|=e^{xt}$, where
\begin{equation}
    x=\frac{\ln|\gb\xi|}{t} \quad \text{is a speed in logarithmic space,}
\end{equation}
so that $|\gb\xi|^{-(q+d)}=e^{-(q+d)xt}$ and
\begin{equation}
    c=\frac1{2\pi i}\int_{\sigma-i\infty}^{\sigma+i\infty}\left[\frac{K}{q+d} +\breve a(q)\right]\phi_{q}(\eta)\,e^{t\Phi(q)}\,dq,
    \quad \Phi(q)=\Lambda(q)-(q+d)x, \label{eq:mellin2}
\end{equation}
which involves the phase  function $\Phi(q)$. The above expression pertains to the uniformly valid solution for $c$. Since $\Lambda(q)$ is a cumulant generating function, it is convex ($\Lambda''(q)>0$). On the real axis, $\Phi$ therefore has a single minimum, at the saddle point $q_s(x)$, defined by the root of
\begin{equation}
    x=\Lambda'(q_s), \label{saddleequation}
\end{equation}
and $q_s$ increases with $x$. The pole at $q=-d$ corresponds to the critical speed
\begin{equation}
    x_*=\Lambda'(-d), \label{eq:xstar}
\end{equation}
at which $q_s=-d$. By convexity $x_*<U=\Lambda'(0)$. Three cases arise: $q_s>-d$ ($x>x_*$: bulk and large-deviation tail, with the Taylor-dispersion regime at $x\approx U$), $q_s<-d$ ($x<x_*$: plateau), and $q_s\to-d$ ($x\approx x_*$: transition zone, where the saddle and the pole merge). We treat the regimes below using the method of steepest descent.

\subsection{Log-normal bulk and its large-deviation tails ($x>x_*$)}
\label{sec:bulk}

For $x>x_*$ the saddle lies to the right of the pole, so the contour can be moved onto the
steepest-descent path $q=q_s+iy$ without crossing a singularity. Expanding the phase about
the saddle,
\begin{equation}
    \Phi(q)\simeq\Phi(q_s)+\tfrac12\Lambda''(q_s)(q-q_s)^2=\Phi(q_s)-\tfrac12\Lambda''(q_s)\,y^2 ,
\end{equation}
and evaluating the slowly varying part of the integrand in~\eqref{eq:mellin2} at $y=0$
(Laplace's method), the remaining Gaussian integral gives
\begin{equation}
    c(\eta,|\gb\xi|,t)\simeq \frac{a(q_s)\,\phi_{q_s}(\eta)}{\sqrt{2\pi\Lambda''(q_s)\,t}}\;e^{t\Phi(q_s)},
    \qquad \Phi(q_s)=\Lambda(q_s)-(q_s+d)\,x .
    \label{eq:tail}
\end{equation}
For $x_*<x<v_{\mathrm{max}}$, this includes the central bulk Gaussian structure together with its large-deviation tails, and it requires the full knowledge of the function $\Lambda(q)$ and its counterpart $\phi_q(\eta)$ and it depends on the function $a(q_s)$, which carries the memory of the early diffusive stage. The expression holds at fixed $x$ with relative error $\mc O(t^{-1})$.

\emph{Large-deviations tail:} First, let us examine the large-deviation tails away from the central bulk region. The exponent of $c$ in~\eqref{eq:tail} is $t\Phi(q_s)$. Writing
\begin{equation}
    \Phi(q_s)=-I(x)-dx \qquad \text{with} \qquad I(x)\equiv q_s(x)\,x-\Lambda(q_s(x)) \label{eq:ratefunction}
\end{equation}
separates the factor
$e^{-dxt}=|\gb\xi|^{-d}$ of the volume element from the remainder $e^{-I(x)t}$,
which governs the mass. Since $q_s=(\Lambda')^{-1}(x)$, $I$ is the Legendre transform of
$\Lambda$, $I=\sup_q[qx-\Lambda(q)]$, known in large-deviation theory as the \textit{rate function} or \textit{entropy function}. We then have 
\begin{equation}
    c\sim e^{-[I(x)+dx]\,t}\qquad (x_*<x<v_{\max}).
    \label{eq:c_exponent}
\end{equation}
The mass per unit $X=\ln|\gb\xi|$ is obtained by integrating $c$ over $\eta$, using
$\int_{\Omega_\eta}\phi_q\,\mathrm d\eta=1$, and multiplying by the volume element
$S_d\,e^{dX}=S_d\,e^{dxt}$ (see Section~\ref{sec:taylor}); the factor $e^{dxt}$ cancels the $e^{-dxt}$ in
$\Phi(q_s)$:
\begin{equation}
    \frac{\mathrm dM}{\mathrm dX}\simeq \frac{S_d\,a(q_s)}{\sqrt{2\pi\Lambda''(q_s)\,t}}\;e^{-I(x)\,t}.
    \label{eq:massdensity}
\end{equation}
The rate function $I$ has the following properties, all consequences of the convexity of
$\Lambda$ with $\Lambda(0)=0$ and $\Lambda'(0)=U$:
\begin{itemize}
  \item $q_s(x)$ increases with $x$, with $q_s=0$ at $x=U$ and $q_s=-d$ at $x=x_*$. Hence
        $q_s\in(-d,0)$ for $x_*<x<U$ (rear tail) and $q_s>0$ for $x>U$ (forward tail).
  \item $I'(x)=q_s(x)$ and $I''(x)=1/\Lambda''(q_s)>0$, so $I$ is convex with its minimum
        $I(U)=0$; therefore $I(x)>0$ for $x\neq U$.
  \item Along the tails the local slopes are $\mathrm d\ln c/\mathrm d\ln|\gb\xi|=-(q_s+d)$
        and $\mathrm d\ln(\mathrm dM/\mathrm dX)/\mathrm dX=-q_s$ (figure~\ref{fig:sch}),
        and the cross-section is $\phi_{q_s}(\eta)$, which varies continuously along the tail.
\end{itemize}

\emph{Central bulk.} For $x-U=\mc O(t^{-1/2})$ one has $q_s=\mc O(t^{-1/2})$, and the
quadratic (Taylor-dispersion) approximation $\Lambda(q)\simeq Uq+Dq^2$ from~\eqref{eq:eigexp}
suffices, with error $\mc O(q^3)=\mc O(t^{-3/2})$ in $\Lambda$, i.e.\ $\mc O(t^{-1/2})$ in the
exponent $t\Phi$. Then $\Phi(q)=Uq+Dq^2-(q+d)x$, and the saddle condition
\eqref{saddleequation} and the phase at the saddle become
\begin{equation}
    q_s=\frac{x-U}{2D}, \qquad  \Phi(q_s)=-dx-\frac{(x-U)^2}{4D},
    \qquad\text{i.e.}\quad I(x)\simeq\frac{(x-U)^2}{4D}.
    \label{eq:ps_bulk}
\end{equation}
Expression~\eqref{eq:tail} then yields
\begin{equation}
    c(\eta,|\gb\xi|,t)\simeq a(q_s)\frac{\phi_{q_s}(\eta)}{\sqrt{4\pi Dt}}
    \exp\left[-dxt-\frac{(x-U)^2}{4D}t\right].
\end{equation}
Since $q_s\approx 0>-d$, the pole is far away and $a(q_s)$ is smooth. Moreover
$\phi_{q_s}\simeq\phi_0=m_\infty$ and $a(q_s)\simeq a(0)$. The amplitude is fixed by mass
conservation: from~\eqref{eq:M_c} and~\eqref{eq:mellin_sol}, $\mc M_0=S_d\,a(0)\,m_\infty$ with
$\Lambda(0)=0$, and $\int_{\Omega_\eta}\mc M_0\,\mathrm d\eta=M_{\rm tot}$ is the total mass, so
$a(0)=M_{\rm tot}/S_d$. Finally, since $e^{-dxt}=|\gb\xi|^{-d}$ and
$(x-U)t=\ln|\gb\xi|-Ut$, we write
\begin{equation}
    c(\eta,|\gb\xi|,t)\simeq \frac{M_{\rm tot}}{S_d}\frac{|\gb\xi|^{-d}m_\infty(\eta)}{\sqrt{4\pi Dt}}
    \exp\left[-\frac{(\ln|\gb\xi|-Ut-X_0)^2}{4Dt}\right]. \label{eq:bulk}
\end{equation}
This is a log-normal profile in $|\gb\xi|$ with the Boltzmann cross-section $m_\infty(\eta)$,
and $\int c\,\mathrm d^d\xi\,\mathrm d\eta=M_{\rm tot}$. The shift $X_0$ comes from the variation
of $a(q)$ across the saddle. Writing
$a(q_s)\simeq a(0)\big[1+\tfrac{a'(0)}{a(0)}\tfrac{x-U}{2D}\big]\simeq a(0)\exp\!\big[\tfrac{a'(0)}{a(0)}\tfrac{x-U}{2D}\big]$
and combining with the Gaussian factor, with $s=\ln|\gb\xi|-Ut$,
\begin{equation} \label{eq:X0}
    -\frac{s^2}{4Dt}+\frac{a'(0)}{a(0)}\frac{s}{2Dt}=-\frac{(s-X_0)^2}{4Dt}+\mc O(t^{-1}),\qquad X_0=\frac{a'(0)}{a(0)}.
\end{equation}
The parameter $X_0$ is $\mc O(1)$ and set by the early diffusive stage, a higher-order
correction relative to the $\mc O(\sqrt t)$ width of the bulk. In terms of $N(\eta,X,t)$,
which obeys the Taylor-dispersion equation~\eqref{eq:log_transport}, the solution can be written as
\begin{equation}
    N(\eta,X,t)\simeq \frac{M_{\rm tot}}{S_d}\frac{m_\infty(\eta)}{\sqrt{4\pi Dt}}
    \exp\left[-\frac{(X-Ut-X_0)^2}{4Dt}\right]. \label{eq:bulkN}
\end{equation}

\subsection{Plateau ($x<x_*$)}\label{sec:plateau}
For $x<x_*$ the saddle point lies on the left of $q=-d$. To pass the contour through it, the original line $\mathrm{Re}\{q+d\}=\sigma>0$ must be moved to the left across the simple pole at $q=-d$. By Cauchy's residue theorem this contributes a residue,
\begin{equation}
    c(\eta,|\gb\xi|,t)=\mathop{\mathrm{Res}}_{q=-d}\Big\{\Big[\frac{K}{q+d}+\breve a(q)\Big]\phi_{q}(\eta)\,e^{t\Phi(q)}\Big\}
    +\frac1{2\pi i}\int_{q_s-i\infty}^{q_s+i\infty}\!\cdots\,dq .
\end{equation}
Only the $K/(q+d)$ term has a pole, so
\begin{equation}\label{eq:residueplat}    \mathop{\mathrm{Res}}_{q=-d}=\lim_{q\to-d}\Big[(q+d)\frac{K}{q+d}\phi_{q}(\eta)\,e^{t[\Lambda(q)-(q+d)x]}\Big]=K\,\phi_{-d}(\eta)\,e^{\Lambda(-d)t}.
\end{equation}
The residue is independent of $x$, since $(q+d)x=0$ at the pole. The remaining integral along $q=q_s+iy$ is evaluated by Laplace's method exactly as in~\eqref{eq:tail}, giving a contribution of order $e^{t\Phi(q_s)}$. Since the phase $\Phi$ is convex with its minimum on the real axis at $q_s<-d$, we have $\Phi(q_s)<\Phi(-d)=\Lambda(-d)$, so the contribution of the remaining integral is exponentially smaller than the residue. (Further poles of the full transform, $q=-d-2,-d-4,\dots$ (see Section~\ref{sec:justify}) and contribute terms of order $e^{\Lambda(-d-2)t+2xt}$, $e^{\Lambda(-d-4)t+4xt},\dots$ and smaller, which are also subdominant.) Retaining only the residue, the leading behaviour of $c$ is given by
\begin{equation}
    c(\eta,t)\simeq K\phi_{-d}(\eta)\,e^{\Lambda(-d)t}, \label{eq:plateau}
\end{equation}
a profile independent of $\gb\xi$ that decays uniformly at the rate $-\Lambda(-d)$, set by the whole cross-section through the eigenvalue of $\LL-dv=\LL+qv$. Note that~\eqref{eq:plateau} is exactly the solution of $c_t=(\LL-dv)c$ that is regular at $\gb\xi=\mathbf 0$ (it is the residue $c_0$ of Section~\ref{sec:justify}, so $K\phi_{-d}(\eta) e^{\Lambda(-d)t}=c_0(\eta,t)$).

\subsection{Transition zone ($\ln|\gb\xi|-x_*t=O(\sqrt t)$)}\label{sec:transition}
As $x\to x_*$ the saddle point $q_s\to-d$ merges with the pole $q=-d$, and the standard Laplace's method fails because the factor $K/(q+d)$ is no longer slowly varying across the saddle. We therefore keep the pole exactly and expand everything else about $q=-d$. Let $s=\ln|\gb\xi|-x_*t=(x-x_*)t$. Since $\Phi(-d)=\Lambda(-d)$, $\Phi'(-d)=\Lambda'(-d)-x=x_*-x$ and $\Phi''(-d)=\Lambda''(-d)$, the phase factor becomes
\begin{equation}
    t\Phi(q)\simeq\Lambda(-d)\,t-(q+d)s+\tfrac12\Lambda''(-d)\,t(q+d)^2 . \label{eq:phase_trans}
\end{equation}
The relevant range of integration for~\eqref{eq:mellin2}  is $q+d=\mc O(t^{-1/2})$, so $\phi_{q}\simeq\phi_{-d}$, and the regular part $\breve a(q)\simeq\breve a(-d)$ contributes only a relative $O(t^{-1/2})$ correction, because $\int e^{\frac12\Lambda''t(q+d)^2}dq\sim t^{-1/2}$ whereas the pole term contribution is $\mc O(1)$. Hence, with $\tau=\Lambda''(-d)\,t$,
\begin{equation}
    c\simeq K\phi_{-d}(\eta)\,e^{\Lambda(-d)t}\,J(s),\quad
J(s)=\frac1{2\pi i}\int_{\sigma-i\infty}^{\sigma+i\infty}\frac1p\,e^{\frac12\tau p^2-ps}\,\mathrm dp,\quad p=q+d .\label{eq:I}
\end{equation}
The integral $J(s)$ in this expression is a standard representation of the complementary error function and therefore
\begin{equation}
    c(\eta,|\gb\xi|,t)\simeq \frac{K}{2} e^{\Lambda(-d)t}\phi_{-d}(\eta)\,
    \mathrm{erfc}\left[\frac{\ln|\gb\xi|-x_*t}{\sqrt{2\Lambda''(-d)\,t}}\right]. \label{eq:transition}
\end{equation}
This is a layer of width $O(\sqrt t)$ travelling at the speed $x_*$. It tends to the plateau~\eqref{eq:plateau} as $s\to-\infty$  and to zero as $s\to+\infty$. In that limit, $\mathrm{erfc}(z)\sim e^{-z^2}/(z\sqrt\pi)$ and the profile becomes $\propto e^{-s^2/2\Lambda''t}$, which matches the large-deviation tail~\eqref{eq:tail} near $x_*$, because $\Phi(q_s)-\Phi(-d)\simeq-(x-x_*)^2/2\Lambda''(-d)$.

\subsection{Justification for neglecting longitudinal diffusion}\label{sec:justify}

Retaining the longitudinal molecular diffusion term $\delta\Delta_{\gb\xi}c$, modifies~\eqref{eq:mellin_eq} to the more general one,
\begin{equation}
    \tilde c_t(q)=[\LL+qv]\tilde c(q)+\delta\,q(q+d-2)\,\tilde c(q-2).
\end{equation}
The diffusion term shifts $q+d$ by $2$. Near $\gb\xi=\mathbf 0$ (which is the stagnant region where longitudinal diffusion can become important) we need to account for the poles at $q+d=0,-2,-4,\dots$. Taking residues of the equation at these poles give
\begin{align}
    \partial_t c_0&=(\LL-dv)c_0+2d\,\delta\,c_2,\\
    \partial_t c_2&=(\LL-(d+2)v)c_2+4(d+2)\,\delta\,c_4,\ \dots
\end{align}
The coupling between different Mellin coefficients runs strictly one-way: each coefficient is forced by the next higher-order state. The unforced component of the second coefficient decays as  $c_2\sim e^{\Lambda(-d-2)t}$, whereas the leading-order plateau state develops at the rate $c_0\sim e^{\Lambda(-d)t}$. By examining the governing eigenvalue problem, the derivative $\Lambda'(q)$ represents a weighted spatial average of the local longitudinal strain rate $v(\eta)$.  Because the fluid accelerates away from the stagnation region, the local velocity is non-negative everywhere $(v\geq 0)$, which implies its weighted average must also be positive $(\Lambda'(q)>0)$. Consequently, the eigenvalue function $\Lambda$ is strictly increasing, guaranteeing that
\begin{equation}
    \Lambda(-d-2)<\Lambda(-d).
\end{equation}
The forcing of $c_0$ by longitudinal diffusion is therefore exponentially small relative to $c_0$ at large times, and the plateau obeys $c_t=(\LL-dv)c$ at leading order. Away from the origin ($|\gb\xi|\gg1$), the ratio of longitudinal diffusion to stretching is
\begin{equation}
    \frac{\delta\Delta_{\gb\xi}c}{v\nabla_{\gb\xi}\cdot(\gb\xi c)}=O(|\gb\xi|^{-2})=O(e^{-2xt}),
\end{equation}
which is exponentially small for any $x>0$. Longitudinal diffusion is therefore negligible in all three regions.

\section{Some structural consequences} \label{sec:structural}

\subsection{Where the scalar mass resides}
\label{sec:mass}

The answer to this question depends on whether one asks about concentration or about mass,
because the two differ by the volume element, which grows exponentially with $X=\ln|\gb\xi|$.

\emph{Concentration.} The plateau concentration~\eqref{eq:plateau} is
$c_{\rm plateau}\simeq K\,\phi_{-d}(\eta)\,e^{\Lambda(-d)t}$, independent of $\gb\xi$. The bulk
concentration~\eqref{eq:bulk} is largest at its centre $X=Ut$, where $|\gb\xi|^{-d}=e^{-dUt}$, so
$c_{\rm bulk}\simeq (M_{\rm tot}/S_d)\,m_\infty(\eta)\,e^{-dUt}/\sqrt{4\pi Dt}$. Ignoring algebraic
factors in $t$ and $\mc O(1)$ constants (denoted by $\sim$ in this section), their ratio is
\begin{equation}
  \frac{c_{\rm plateau}}{c_{\rm bulk}}\sim e^{\,[\Lambda(-d)+dU]\,t}.
  \label{eq:c-ratio}
\end{equation}
The exponent is positive: for $D>0$, $\Lambda$ is strictly convex with $\Lambda(0)=0$ and
$\Lambda'(0)=U$, so $\Lambda(q)>Uq$ for $q\neq0$ (Section~\ref{sec:Lambda}), and in particular
$\Lambda(-d)>-dU$. Moreover, by~\eqref{eq:c_exponent} the exponent of the concentration along the
tails is $-[I(x)+dx]$, whose derivative with respect to $x$ is $-(q_s+d)<0$ for $x>x_*$. The
concentration therefore decreases monotonically from the plateau outwards, and the plateau is the
global maximum of the concentration, exceeding the peak bulk concentration by the exponentially
large factor~\eqref{eq:c-ratio}.

\emph{Mass.} The mass in a slice of width $\mathrm dX$ is
\begin{equation}
  \mathrm dM=S_d\,e^{dX}\Big(\int_{\Omega_\eta} c\,\mathrm d\eta\Big)\mathrm dX ,
  \label{eq:slice}
\end{equation}
so the question is whether the plateau's higher concentration can compensate for its smaller volume
element. Between the upper edge of the plateau, $X=x_*t$, and the centre of the bulk, $X=Ut$, the ratio
of the volume elements is
\begin{equation}
  \frac{e^{dx_*t}}{e^{dUt}}=e^{-d\,(U-x_*)\,t}.
  \label{eq:vol-ratio}
\end{equation}
Integrating~\eqref{eq:slice} over the plateau, with $\int_{\Omega_\eta}\phi_{-d}\,\mathrm d\eta=1$,
\begin{equation}
  M_{\rm plateau}=S_d\int_{-\infty}^{x_*t}e^{dX}\,\mathrm dX\;K\,e^{\Lambda(-d)t}
  =\frac{S_d K}{d}\,e^{\,[\Lambda(-d)+d\,x_*]\,t}.
  \label{eq:mass-plateau}
\end{equation}
Since the bulk carries essentially all the mass, $M_{\rm bulk}\simeq M_{\rm tot}$, the ratio
$M_{\rm plateau}/M_{\rm bulk}$ is the product of~\eqref{eq:c-ratio} and~\eqref{eq:vol-ratio}, as it
should be, and the definition~\eqref{eq:ratefunction} of $I$ at $x=x_*$ ($q_s=-d$) gives
\begin{equation}
  \frac{M_{\rm plateau}}{M_{\rm tot}}\sim e^{\,[\Lambda(-d)+d\,x_*]\,t}=e^{-I(x_*)\,t}.
  \label{eq:mass-ratio}
\end{equation}
The exponent is negative: $\Lambda$ is strictly convex, so its tangent at $q=-d$, of slope
$\Lambda'(-d)=x_*$, lies below the curve, and evaluating the tangent at $q=0$ gives
$0=\Lambda(0)>\Lambda(-d)+d\,x_*$. Equivalently, the volume deficit~\eqref{eq:vol-ratio} exceeds the
concentration excess~\eqref{eq:c-ratio}. The plateau therefore contains an exponentially small fraction
of the scalar mass.

The same conclusion holds for the transition zone, which joins the plateau to the rear tail. For $x<x_*$
the mass density is $S_d\,e^{dX}c_{\rm plateau}\sim e^{[\Lambda(-d)+dx]t}=e^{-[I(x_*)+d(x_*-x)]t}$. Its
exponent is the tangent line to $I$ at $x_*$ (slope $q_s=-d$). By convexity this tangent lies below $I$, so the plateau mass density exceeds the continuation $e^{-I(x)t}$ of the tail formula at every $x<x_*$. This is consistent with the residue~\eqref{eq:residueplat} dominating the saddle-point contribution there. At $x=x_*$ the two exponents and their
slopes with respect to $x$ coincide (the slope of the logarithm of the mass density is $-q_s=d$ on both
sides), so the exponent is continuous with continuous slope across the transition zone. Because $d(x_*-x)>0$, the plateau mass density is nevertheless smaller than $e^{-I(x_*)t}$ for all $x<x_*$, and the total plateau carries only a fraction $e^{-I(x_*)t}$ of the scalar mass.

\emph{Summary.} The mass density and the concentration differ only by the factor $e^{dxt}$, which turns
the exponent $I(x)+dx$ of the concentration into the exponent $I(x)$ of the mass density. This is what
moves the maximum from the plateau, where $I(x)+dx$ is smallest, to the bulk, where $I(x)$ is smallest.
A probe near the stagnation point therefore records the plateau, whereas the bulk, at a distance
$(U-x_*)t$ in $\ln|\gb\xi|$, holds essentially all of the scalar.

\subsection{Collapse onto a single Gaussian for uniform strain rate}
\label{sec:uniform}

Let $v$ be independent of $\eta$, as in the Burgers vortex of Section~\ref{sec:burgers} (there $v=2$, $d=1$, $\delta=1$).
Since $\mathcal{L}m_\infty=0$, we have $(\mathcal{L}+qv)\,m_\infty=qv\,m_\infty$ for every $q$, so
\begin{equation}
  \phi_q=m_\infty,\qquad \Lambda(q)=vq\quad\text{(exactly)},\qquad
  U=x_*=v,\qquad D=\tfrac12\Lambda''=0 .
  \label{eq:uniform-params}
\end{equation}
The phase function~\eqref{eq:mellin2} is then linear in $p$, $\Phi(p)=-dv+p\,(v-x)$, so $\Phi'=v-x$
is independent of $p$ and there is no saddle point for any $x$ (the stationary-phase
condition degenerates at $x=v$). The steepest-descent analysis is therefore not needed
here; instead the contour is simply closed to the left or right of the pole according
to the sign of $x-v$, or the Mellin
transform~\eqref{eq:mellin} is evaluated exactly,
\begin{equation}
  c(\eta,|\boldsymbol\xi|,t)=e^{-dvt}\,m_\infty(\eta)\,A\!\left(|\boldsymbol\xi|e^{-vt}\right),
  \qquad
  A(\zeta)=\frac{1}{2\pi i}\int_{\sigma-i\infty}^{\sigma+i\infty}a(q)\,\zeta^{-(q+d)}\,\mathrm{d}q ,
  \label{eq:uniform-exact}
\end{equation}
that is, a fixed profile $A$, set by the early diffusive stage, that is stretched by $e^{vt}$ and diluted by $e^{-dvt}$. Setting $\boldsymbol\xi=0$ gives
$K=A(0)$, the residue of $a$ at $q=-d$.

For a uniform strain the full problem, including diffusion, is solved
exactly by a Gaussian profile~\cite{townsend1951diffusion}, 
\begin{equation}
  c=\frac{M_{\rm tot}\,m_\infty(\eta)}{(2\pi\Sigma^2)^{d/2}}
  \exp\!\left(-\frac{|\boldsymbol\xi|^2}{2\Sigma^2}\right)
  \;\longrightarrow\;
  \frac{M_{\rm tot}\,m_\infty(\eta)}{(2\pi\Sigma_\infty^2)^{d/2}}\,e^{-dvt}
  \exp\!\left(-\frac{|\boldsymbol\xi|^2e^{-2vt}}{2\Sigma_\infty^2}\right)
  \quad(vt\gg1).
  \label{eq:uniform-gauss}
\end{equation}
where the variance obeys
$\dot\Sigma^2=2v\Sigma^2+2\delta$, so
$\Sigma^2(t)=\Sigma_\infty^2e^{2vt}-\delta/v$ with $\Sigma_\infty^2=\Sigma_0^2+\delta/v$ ($\Sigma_0$ is the initial width;
$\Sigma_0=0$ for a point release). The above limiting expression is \eqref{eq:uniform-exact} with a Gaussian $A(\zeta)$, and it also confirms that
the neglected longitudinal diffusion only produces the exponentially small
correction $-\delta/v$ to $\Sigma^2$ (although it enters indirectly  through $\Sigma_\infty$). Comparison with \eqref{eq:uniform-exact} gives
\begin{align}
&A(\zeta)=  \frac{M_{\rm tot}}{(2\pi\Sigma_\infty^2)^{d/2}}
  \exp\!\left(-\frac{\zeta^2}{2\Sigma_\infty^2}\right), \qquad \zeta = |\gb\xi|e^{-v t},\qquad K=A(0)=\frac{M_{\rm tot}}{(2\pi\Sigma_\infty^2)^{d/2}},\\
  &X_0=\frac{a'(0)}{a(0)}=\langle\ln|\boldsymbol\xi|\rangle
  =\ln\Sigma_\infty+\tfrac12\left[\psi(d/2)+\ln 2\right],
\end{align}
where $\psi$ is the digamma function (for $d=1$, $X_0=\ln\Sigma_\infty-\tfrac12(\gamma+\ln2)$). For the Burgers vortex $m_\infty\propto e^{-r^2/2}$ is itself a Gaussian in $r$, so the
asymptotic solution is a fully Gaussian cloud: of fixed radial thickness $O(1)$ and
of an evolving axial width $\Sigma\sim e^{vt}$.

In terms of $X=\ln|\boldsymbol\xi|$ and $N=|\boldsymbol\xi|^dc$, \eqref{eq:uniform-exact} reads
\begin{equation}
  N=m_\infty(\eta)\,B(X-vt),\qquad B(s)=e^{ds}A(e^{s}),
\end{equation}
so the cloud is translated at speed $v$ in logarithmic space without spreading, which is the
$D=0$ limit of~\eqref{eq:bulkN}. The Gaussian of width $\sqrt{4Dt}$ in $X$ is replaced by the fixed $O(1)$ profile $B$ from the early shape. The three regions become
\begin{itemize}
  \item \emph{Bulk.} The mass is carried by $|\boldsymbol\xi|\sim e^{vt}$, with the Boltzmann cross-section $m_\infty$.
  \item \emph{Plateau.} For $|\boldsymbol\xi|\ll e^{vt}$, $c\to K\,m_\infty\,e^{-dvt}$, which is~\eqref{eq:plateau} with
  $\phi_{-d}=m_\infty$ and $\Lambda(-d)=-dv$. It is the core of the Gaussian and decays at the rate $dv$.
  \item \emph{Transition zone.} Here $\tau=\Lambda''(-d)\,t=0$, so the $\mathrm{erfc}$ layer~\eqref{eq:transition}
  degenerates into a step at $\ln|\boldsymbol\xi|=x_*t=vt$, and its width is no longer $O(\sqrt t)$. It is the $O(1)$ shoulder of $B$,
  which was dropped in Section~\ref{sec:transition} as the relative $O(t^{-1/2})$ regular part $\breve a$ and is now the same size as the pole term.
\end{itemize}
Since $U-x_*=0$, the separation $(U-x_*)t$ between bulk and plateau vanishes, and
since $D=0$ the widths of all three regions are $O(1)$ rather than $O(\sqrt t)$. All three
therefore sit at $\ln|\boldsymbol\xi|=vt+O(1)$ and merge into the single Gaussian
\eqref{eq:uniform-gauss}. The plateau mass $\sim e^{-d^2Dt}$ tends to $O(1)$, so the plateau is no longer a
vanishing fraction of the mass. 
For the Burgers vortex ($v=2$), $\Sigma^2=(\Sigma_0^2+\tfrac12)e^{4t}-\tfrac12$ and the amplitude decays as $e^{-2t}$.

\subsection{How accurate is the bulk Gaussian?}\label{sec:gauss}

The Taylor-dispersion description of the bulk, $\Lambda\simeq Uq+Dq^2$, keeps only the
first two terms of a series, so it is worth asking when they suffice. The relevant time
scale is the transverse relaxation time $1/\lambda_1$, where $\lambda_1$ is the smallest
non-zero eigenvalue of $\mathcal L$. It is the rate at which the
cross-section of the cloud relaxes to $m_\infty(\eta)$. Seen along a solute molecule,
$\mathrm d\ln|\boldsymbol\xi|/\mathrm dt=v(\eta(t))$, where $\eta(t)$ is a stationary
random process that decorrelates over a time $1/\lambda_1$. The log-stretch is thus a sum
of correlated increments, and it becomes Gaussian only after many of them have been
accumulated. The strain gained across the section during one relaxation time,
\begin{equation}
    \Pi = \frac{v_{\max}-v_{\min}}{\lambda_1},
\end{equation}
measures how far each increment departs from the
mean, and hence how far the cloud is from Gaussian. Three conditions control how clearly
the Gaussian is seen:
\begin{enumerate}
\item the cross-section has relaxed, $t\gg 1/\lambda_1$;
\item the log-normal spreading is resolved: the width $\sqrt{2Dt}$ of the bulk must
exceed the $O(1)$ width set by the early diffusive stage (the profile $B$ of
Section~\ref{sec:uniform}), that is, $t\gg 1/2D$; similarly, the separation
$(U-x_*)t\simeq 2dDt$ of bulk and plateau must exceed unity;
\item the skewness of $\ln|\boldsymbol\xi|$, which is of order $\Lambda_3/(D^{3/2}t^{1/2})$
with $\Lambda_3$ the third Taylor coefficient of $\Lambda$, must be small.
\end{enumerate}
For $\Pi\ll1$ (nearly uniform straining), $D\sim\Pi^2\lambda_1$, so condition~2 is the binding one, requiring
$t\gg1/(\Pi^2\lambda_1)\gg1/\lambda_1$, and the skewness is of order $(\lambda_1t)^{-1/2}$,
so condition~3 follows from condition~1. The quadratic closure is then accurate even
at $q=-d$ (Appendix~\ref{app:taylor}), giving $x_*\simeq U-2dD$, $\Lambda(-d)\simeq-dU+d^2D$ and
$I(x)\simeq(x-U)^2/4D$. The price is that all five regions are close together. Until
$t\gg1/2D$ the cloud is indistinguishable from the uniform-strain Gaussian of
Section~\ref{sec:uniform}. The large-P\'eclet-number tube of Appendix~\ref{app:largepe} is an example, with
$D=72\pi/\Pec^3$ and the structure emerging only for $t\sim\Pec^3$. For
$\Pi\gtrsim1$, the strain varies appreciably within one relaxation time. The centre of the
cloud then becomes Gaussian only after times longer than $1/\lambda_1$ and the tails and plateau cannot be obtained from $U$ and $D$ alone,
because the quadratic closure fails.

\section{The principal eigenvalue and eigenfunction}\label{sec:Lambda}

As established in the two preceding sections, the principal eigenvalue $\Lambda(q)$ and eigenfunction $\phi_q(\eta)$ of the operator $\LL+qv$ control the long-time asymptotics. Unlike classical Taylor dispersion, where the quadratic closure $\Lambda\approx Uq+Dq^2$ is sufficient, here we need $\Lambda$ at $q=-d$ (plateau and transition zone) and at the saddle point $q_s(x)$ (large-deviation tail), which can be far from $q=0$. The eigenvalue problem is
\begin{equation}
  (\LL+qv)\,\phi_q=\Lambda(q)\,\phi_q,\qquad \int_{\Omega_\eta}\phi_q\,\mathrm{d}\eta=1,
  \label{eq:eig}
\end{equation}
with vanishing total flux, $J=0$, on the transverse boundaries (the Robin condition~\eqref{eq:robin} for the tube and $c_\eta=0$ at the wall for the surface flows). In this section we examine the properties of $\Lambda(q)$ and $\phi_q$ that are needed to interpret the results and explain why the von K\'arm\'an flow behaves differently.

\subsection{Self-adjoint form, variational principle and exact bounds}

Since $m_\infty'=-\beta m_\infty$, writing $\phi_q=m_\infty h_q$ gives $\phi_q'+\beta\phi_q=m_\infty h_q'$, and \eqref{eq:eig} becomes
\begin{equation}
  \frac{1}{m_\infty}(\alpha m_\infty h_q')'+q\,v\,h_q=\Lambda\,h_q,\qquad h_q'=0\ \text{on}\ \partial{\Omega_\eta} .
  \label{eq:eig_h}
\end{equation}
The flux condition becomes a plain Neumann condition on $h_q$, and the operator is self-adjoint in the weighted inner product $\langle f,g\rangle=\int_{\Omega_\eta} m_\infty fg\,\mathrm{d}\eta$. The function $h_q$ is also the eigenfunction of the adjoint operator $\LL^\dagger+qv$. Hence
\begin{equation}
  \Lambda(q)=\sup_{h}\ \frac{q\int_{\Omega_\eta} v\,m_\infty h^2\,\mathrm{d}\eta-\int_{\Omega_\eta} \alpha\, m_\infty h'^2\,\mathrm{d}\eta}{\int_{\Omega_\eta} m_\infty h^2\,\mathrm{d}\eta},
  \qquad
  \Lambda'(q)=\frac{\int_{\Omega_\eta} v\,m_\infty h_q^2\,\mathrm{d}\eta}{\int_{\Omega_\eta} m_\infty h_q^2\,\mathrm{d}\eta}.
  \label{eq:variational}
\end{equation}
The second expression (due to Hellmann--Feynman theorem) shows that $\Lambda'(q)$ is a weighted mean of the local strain rate $v$, with the weight $m_\infty h_q^2$. Several exact properties follow:
\begin{itemize}
  \item $\Lambda$ is a supremum of functions linear in $q$, so it is convex, and $\Lambda''>0$ unless $v$ is uniform (i.e., $\Lambda(q)=\sup_{h} (A_hq-B_h)$ sits as the upper envelope at each $q$ covering an infinite collection of straight lines). Moreover $v_{\min}<\Lambda'(q)<v_{\max}$ (on the discrete branch), so the rate function $I(x)=\sup_q[qx-\Lambda(q)]$ is finite on $(v_{\min},v_{\max})$ and infinite outside $[v_{\min},v_{\max}]$. The Taylor-dispersion closure $\Lambda(q)=Uq+Dq^2$, which has constant curvature, breaks down for large $|q|$ once it exceeds the physical bound.
  \item The trial function $h=1$ gives $\Lambda(q)\ge Uq$ for all $q$.
  \item Because $\LL\le 0$ with $\LL m_\infty=0$, we have $\Lambda(q)\le q\,v_{\max}$ for $q>0$ and $\Lambda(q)\le q\,v_{\min}$ for $q<0$.
\end{itemize}
At $q=-d$ these give
\begin{equation}
  -dU\ \le\ \Lambda(-d)\ \le\ -d\,v_{\min}.
  \label{eq:bounds}
\end{equation}
For all the flows (besides Burgers vortex) considered here $v_{\min}=0$ and hence $0\le -\Lambda(-d)\le dU$ (Table~\ref{tab:Lambda}) and the plateau decays no faster than the bulk amplitude $e^{-dUt}$ . The same bounds give $x_*=\Lambda'(-d)<U$. The rate function also has a direct meaning: the plateau mass $e^{\Lambda(-d)t+dx_*t}=e^{-I(x_*)t}$ is the cost of accumulating an atypically small log-stretch $x_*t$.

While the exact behaviour of  $\Lambda(q)$ is governed by the full variational principle, its local behaviour can be evaluated via a systematic Taylor expansion about $q=0$, which is discussed in Appendix~\ref{app:taylor}.

\begin{figure}[h!]
\centering
\includegraphics[scale=0.55]{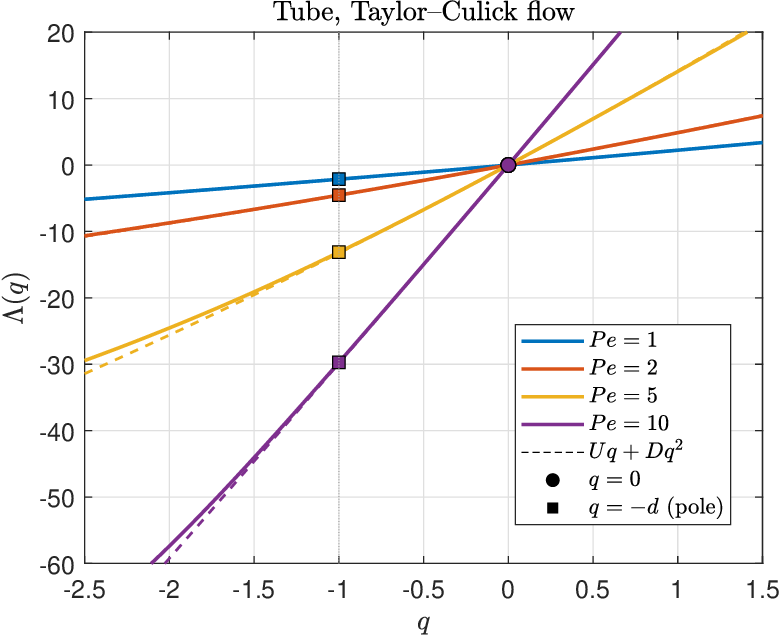}
\includegraphics[scale=0.55]{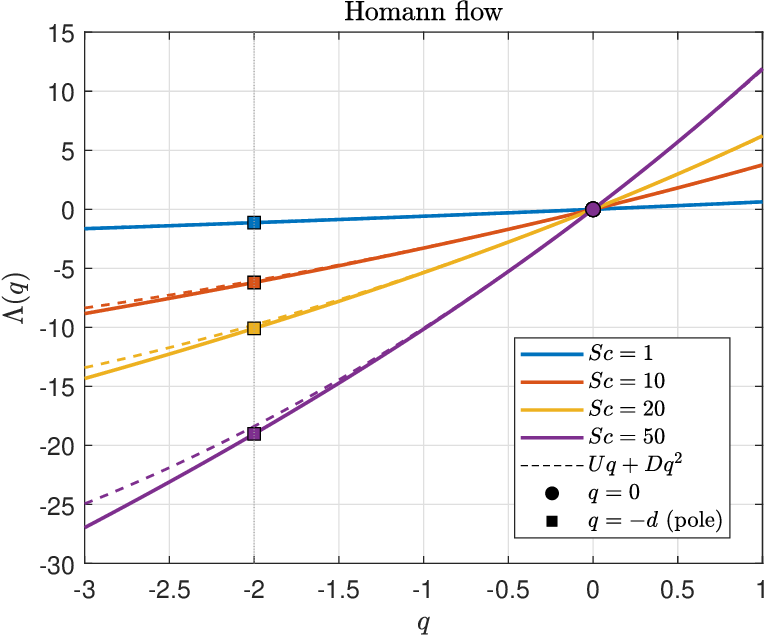}
\includegraphics[scale=0.55]{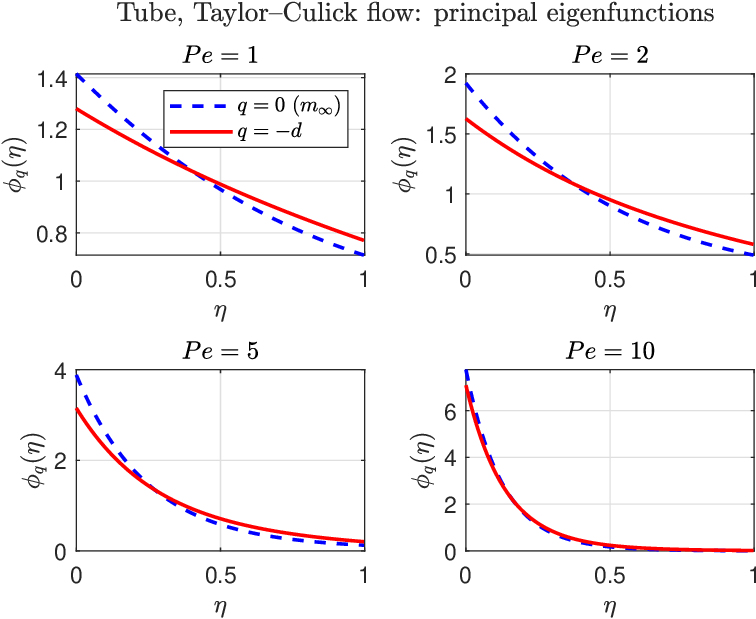}
\includegraphics[scale=0.55]{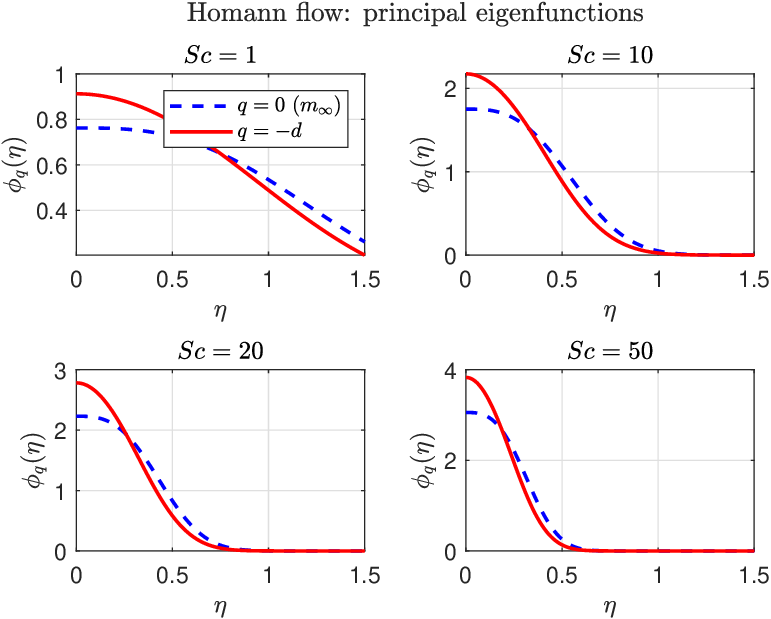}
\caption{Principal eigenvalue and eigenfunction of $\LL+qv$ for the
Taylor--Culick tube flow (left; $\Pec=1,2,5,10$) and the Homann flow
(right; $\Sch=1,10,20,50$). Upper panels: $\Lambda(q)$ (solid) and the
quadratic closure $Uq+Dq^2$ (dashed); circles mark $q=0$ and squares mark
$q=-d$, whose ordinate $\Lambda(-d)$ is the plateau decay rate. Lower panels: the normalised eigenfunction
$\phi_{-d}(\eta)$ (solid) compared with the Boltzmann profile
$\phi_0=m_\infty(\eta)$ (dashed), with $\int\phi_q\,\mathrm{d}\eta=1$.}
\label{fig:eigen}
\end{figure}

Figure~\ref{fig:eigen} shows $\Lambda(q)$ and $\phi_q$ for the Taylor--Culick
tube flow at $\Pec=1,2,5,10$ and for the Homann flow at
$\Sch=1,10,20,50$. In the upper panels, $\Lambda(q)$ is convex with
$\Lambda(0)=0$ and slope $U$ at $q=0$. It agrees with the second-order
(quadratic) closure $Uq+Dq^2$ near $q=0$ and departs from it as $|q|$ grows.
The departure at $q=-d$ is what matters for the plateau, and it grows with
$\Pec$ and $\Sch$. For the Homann flow, $-dU+d^2D=-1.127$ against
$\Lambda(-d)=-1.133$ at $\Sch=1$, but $-18.383$ against $-19.023$ at
$\Sch=50$. The quadratic closure therefore estimates the plateau decay
rate only at small $\Pec$ and $\Sch$, and the full $\Lambda(q)$ is
needed beyond that. Table~\ref{tab:Lambda} gives the numerical values of $U$,
$D$, $\Lambda(-d)$ and $x_*$ together with the quantities $\lambda_1$, and $\Pi$
that control the Gaussianity of the log-stretch (Section~\ref{sec:gauss}).

\begin{table}
  \centering
  \begin{tabular}{lccccccc}
  \toprule
    flow & $U$ & $D$ & $\lambda_1$ & $\Pi$ &  $-dU$ & $\Lambda(-d)$ & $x_*$\\[2pt]\midrule
       Taylor--Culick, Pe$=1$  & 2.1843  & 0.0481 & 14.6786 & 0.214  & $-2.184$  & $-2.1359$  & 2.087\\ % TODO lambda1, Pi, R_K
   Taylor--Culick, Pe$=2$  & 4.7125  & 0.1678 & 15.0191 & 0.418 & $-4.713$  & $-4.5375$  & 4.357\\
    Taylor--Culick, Pe$=5$  & 13.7185 & 0.4608 & 18.0633 & 0.870  & $-13.719$ & $-13.1295$ & 12.399\\
     Taylor--Culick, Pe$=10$ & 30.0988 & 0.2456 & 29.0141 & 1.083  & $-30.099$ & $-29.7312$ & 29.178\\
    \hline
    Homann, Sc$=1$  & 0.6086  & 0.0225 & 2.8365 & 0.353 &  $-1.217$  & $-1.1326$ & 0.527\\
    Homann, Sc$=10$ & 3.5046  & 0.2395 & 16.6733  & 0.600 & $-7.009$  & $-6.1985$ & 2.761\\
    Homann, Sc$=20$ & 5.7540  & 0.4276 & 27.3940  & 0.730 & $-11.508$ & $-10.0863$ & 4.461\\
    Homann, Sc$=50$ & 10.9461 & 0.8774 & 52.1108  & 0.959 & $-21.892$ & $-19.0234$ & 8.356 \\
    \bottomrule
  \end{tabular}
  \caption{Quantities controlling the log-stretch statistics. $U=\Lambda'(0)$
  and $D=\Lambda''(0)/2$ are the mean log-stretching rate and the log-space
  diffusivity, $\lambda_1$ is the first non-zero relaxation rate of
  $\mathcal{L}$, and $\Pi=(v_{\max}-v_{\min})/\lambda_1$.  The last three columns illustrate the bounds \eqref{eq:bounds}
  and the critical speed $x_*=\Lambda'(-d)$.}
  \label{tab:Lambda}
\end{table}

\subsection{The eigenfunction as a spatially evolving cross-section}

Unlike classical Taylor dispersion, where the cloud retains a uniform cross-section, the cross-section here depends on the speed $x=\ln|\gb\xi|/t$. At the centre of the bulk ($x=U$) it is the Boltzmann profile $m_\infty(\eta)$, shown dashed in the lower panels of figure~\ref{fig:eigen}. Moving through the rear tail towards the plateau ($x\to x_*$), it tends to $\phi_{-d}=m_\infty h_{-d}$ (solid). The factor $h_{-d}$ weights regions of weak
strain more heavily, so $\phi_{-d}$ is displaced towards the wall (and, in the tube, away from the axis) relative to $m_\infty$. This displacement is visible at small $\Pec$ and $\Sch$, and on the scale of the plot it is small at large $\Pec$ and $\Sch$, where both profiles are confined to a thin layer. For $q>0$ the displacement is in the opposite direction. The structural distortion of the cross-section explains why the plateau’s decay rate is determined by an integrated average across the entire cross-section rather than just the maximum centerline strain rate. Consequently, because the cloud's cross-sectional profile dynamically shifts from front to back, a probe at the stagnation point fails to represent the true mass distribution of the moving bulk.

\subsection{Why the von K\'arm\'an flow is different}

For Homann and von K\'arm\'an flows, $\alpha=1$, $\beta=2\Sch f$ and $v=\Sch f'$. Let us rewrite \eqref{eq:eig} into a Schr\"odinger problem by introducing $h_q=u(\eta)\exp(\Sch\int_0^\eta f(s)\,\mathrm{d}s)$, so that
\begin{equation}
  u''-V_q(\eta)\,u=\Lambda\,u,\qquad u'(0)=0,\qquad V_q=\Sch^2f^2-\Sch\,(1+q)\,f' ,
  \label{eq:schrodinger}
\end{equation}
where the boundary condition follows from $f(0)=0$; moreover $u(\eta)$ is to be bounded as $\eta\to\infty$. The eigenvalue $\Lambda(q)$ is the top of the spectrum of $\partial_\eta^2-V_q$ on the half-line. Two situations arise:
\begin{itemize}
  \item \emph{Homann flow.} Here $f\sim\eta-\delta^*$ at large $\eta$, so that $V_q\sim\Sch^2\eta^2$ is a confining potential. The spectrum is purely discrete, the principal eigenvalue is simple for every real $q$, and $\Lambda(q)$ is analytic. The tube has a bounded domain and is discrete for the same reason.
  \item \emph{von K\'arm\'an flow.} Here $f\to f_\infty$ and $f'\to0$ exponentially, so that $V_q\to V_\infty=\Sch^2f_\infty^2$ for \emph{every} $q$. The basic spectrum is continuous, $(-\infty,-\Sch^2f_\infty^2]$, and does not depend on $q$; $q$ enters only through the localised term $-\Sch(1+q)f'$. An isolated eigenvalue above the edge exists only if the well is deep enough, so
  \begin{equation}
    \Lambda(q)=\max\{-\Sch^2f_\infty^2,\ \Lambda_d(q)\},
    \label{eq:Lamkarman}
  \end{equation}
  where the discrete branch $\Lambda_d(q)$ exists for $q>q_{c,*}$ and merges into the edge at $q=q_{c,*}$.
\end{itemize}
Since $\Lambda(0)=0>-\Sch^2f_\infty^2$ and $\Lambda'\ge0$ while the edge is independent of $q$,
a bound state at some $q$ implies one at every larger $q$. The bound region is therefore an
interval $(q_{c,*},\infty)$ containing $q=0$, and it suffices to test the most negative value
required by the leading-order structure, $q=-d$. Whether $q=-d$ lies on the discrete branch can be decided from \eqref{eq:schrodinger}. In one dimension (here a half-line with a Neumann condition) a potential with $\int_0^\infty(V_q-V_\infty)\,\mathrm{d}\eta<0$ always has a bound state. Using $\int_0^\infty f'\,\mathrm{d}\eta=f_\infty$, this condition reads
\begin{equation}
  q>q_c\equiv -1-\frac{\Sch\,\mc I}{f_\infty},\qquad \mc I=\int_0^\infty\left(f_\infty^2-f^2\right)\mathrm{d}\eta .
  \label{eq:qs}
\end{equation}
It is sufficient but not necessary, so the true threshold satisfies $q_{c,*}\le q_c$. For $q$ below the threshold, $\Lambda$ sits on the edge and is independent of $q$; consequently $\Lambda'=0$ there.

\begin{figure}
  \centering
  \includegraphics[scale=0.6]{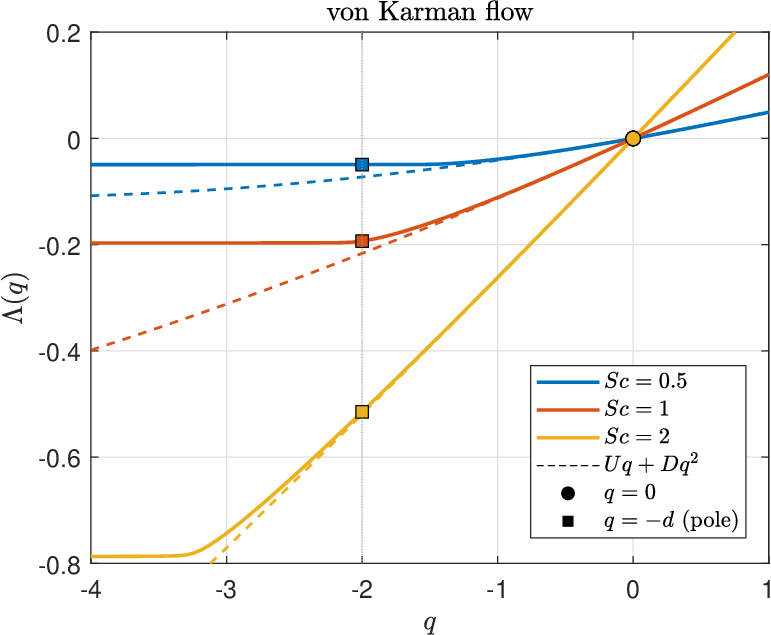}
  \caption{Principal eigenvalue $\Lambda(q)$ for the von K\'arm\'an flow at $\Sch=0.5,1,2$ (solid) and the quadratic closure $Uq+Dq^2$ (dashed). Circles mark $q=0$ and squares mark $q=-d=-2$. Each curve is flat, at the continuum edge $-\Sch^2f_\infty^2$, for $q$ below the threshold $q_{c,*}$ at which the bound state disappears. For $\Sch=2$ the threshold lies beyond the pole ($q_{c,*}\approx-3.2$), for $\Sch=1$ it lies just beyond the pole, and for $\Sch=0.5$ it lies above $-d$, so the plateau decay rate is the edge value.}
  \label{fig:karman}
\end{figure}

For the von K\'arm\'an flow $f_\infty=0.4422$ and $\mc I=0.4901$, and Table~\ref{tab:karman} gives the results for $\Sch=0.5,1,2$. Three cases occur at $q=-d=-2$, and they are visible in figure~\ref{fig:karman}.
\begin{itemize}
  \item $\Sch=2$: $q_c=-3.217<-d$, so a bound state exists at $q=-d$. It is strongly bound (the binding energy $E_b=\Lambda(-d)+\Sch^2f_\infty^2=0.27$, so $u$ decays over $\eta\sim2$), and the plateau behaves as in the confining flows, with $x_*=0.246<U$. The curve $\Lambda(q)$ becomes flat only for $q<q_{c,*}\approx-3.2$, beyond the pole.
  \item $\Sch=1$: $q_c=-2.108$ is just below $-d$, so a bound state exists, but it is only weakly bound. The computed $\Lambda(-2)=-0.1932$ lies just above the edge $-\Sch^2f_\infty^2=-0.1956$, with $E_b=2.4\times10^{-3}$, and $\Lambda$ is flat for $q$ below about $-2.1$, immediately beyond the pole.
  \item $\Sch=0.5$: $q_c=-1.554>-d$, so criterion \eqref{eq:qs} does not apply at $q=-d$ (it requires $\Sch>f_\infty/\mc I\approx0.90$), and the numerics show no bound state: $E_b$ is zero to within truncation error ($-2\times10^{-4}$). The plateau decay rate is then the edge itself, $-\Lambda(-d)=\Sch^2f_\infty^2=0.0489$, and since $\Lambda$ is flat there, $x_*=\Lambda'(-d)=0$. The threshold lies between $-2$ and $-1.55$. In this case there is no normalisable $\phi_{-d}$, and the plateau is carried by the continuum, which decays like $e^{-\Sch^2f_\infty^2 t}$ times a power of $t$.
\end{itemize}

For $\Sch=1$ the small binding energy has three consequences. First, the eigenfunction $u$ decays as $e^{-\sqrt{E_b}\,\eta}$, so $\phi_{-d}$ extends over $\eta\sim20$, far beyond the nominal viscous layer. Second, the bound-state contribution dominates the continuum only for $t\gtrsim1/E_b$, of order $10^{2}$--$10^{3}$ in our units, so at moderate times the plateau decay rate appears to approach the edge $\Sch^2f_\infty^2$ and not the true limit $-\Lambda(-d)$. Third, since $E_b\propto(q-q_{c,*})^2$ as $q\downarrow q_{c,*}$, $\Lambda$ is $C^1$ but has a jump in $\Lambda''$ at $q_{c,*}$, a non-analyticity that does not occur for confining flows. The same applies to the other quantities: for $\Sch=0.5$ the large-deviation structure holds for $q_s>q_{c,*}$ only, so the plateau and the transition zone are controlled by the edge and not by a pole of the discrete branch.

\begin{table}
  \centering
  \begin{tabular}{lcccccccccc}
  \toprule
    $\Sch$ & $U$ & $D$ & $\lambda_1$ & $\Pi$ & $-dU$ & $-\Sch^2f_\infty^2$ & $\Lambda(-d)$ & $x_*$ & $E_b$ & $q_c$\\[2pt]\midrule
    0.5 & 0.0456 & 0.0047 & 0.0492 & 1.84 & $-0.0912$ & $-0.0489$ & $-0.0489^{\dagger}$ & $0^{\dagger}$ & --- & $-1.554$\\
    1   & 0.1168 & 0.0043 & 0.1967 & 0.92 & $-0.2335$ & $-0.1956$ & $-0.1932$ & 0.0365 & $2.4\times10^{-3}$ & $-2.108$\\
    2   & 0.2647 & 0.0026 & 0.7809 & 0.46 & $-0.5294$ & $-0.7823$ & $-0.5146$ & 0.2459 & 0.268 & $-3.217$ \\
    \bottomrule
  \end{tabular}
  \caption{von K\'arm\'an flow ($f_\infty=0.4422$, $f''(0)=0.5102$, $g'(0)=-0.6159$). $\lambda_1$ is the gap between the zero eigenvalue and the continuum edge of $\mathcal{L}$, $\Pi=(v_{\max}-v_{\min})/\lambda_1$, $E_b=\Lambda(-d)+\Sch^2f_\infty^2$ is the binding energy of the discrete branch at $q=-d$, and $q_c$ is the sufficient threshold \eqref{eq:qs}. $^{\dagger}$No bound state exists at $q=-d$ for $\Sch=0.5$: $\Lambda(-d)$ is the continuum edge and $x_*=0$.}
  \label{tab:karman}
\end{table}

\section{Numerical validation}
\label{sec:numerics}

To test the asymptotic structure we solve~\eqref{eq:chat} numerically for
$\hat c(\eta,Y,t)$ with $Y=\ln k$, imposing zero total flux, $J=0$, on the
transverse boundaries (the Robin condition~\eqref{eq:robin} for the tube).
Only the Taylor--Culick tube flow ($d=1$, $\delta=1$, $f=\tfrac12\sin(\pi\eta/2)$) at $\Pec=1$ and $10$ is
considered here. In the variable $Y$ the stretching appears as a translation,
so no grid growing like $e^{vt}$ is required.

\subsection{Numerical method}

The scalar is released as a Gaussian of width $\Sigma_0$ in $z$, uniform in
$\eta$ and normalised so that $M_{\rm tot}=1$, so that $\hat c(\eta,k,0)=e^{-\Sigma_0^2k^2/2}$; we use
$\Sigma_0=3$ for $\Pec=1$ and $\Sigma_0=1$ for $\Pec=10$. The equation is
advanced by a second-order operator splitting with the following ingredients.
\begin{enumerate}
\item \emph{Transverse operator.} $\LL$ is discretised by a conservative
  finite-volume scheme on a grid uniform in $r=\sqrt{\eta}$ (hence clustered
  near the axis), with Scharfetter--Gummel (exponentially fitted) fluxes and
  cell-averaged strain $v=\Delta(\text{stream function})/\Delta\eta$. Mass
  conservation and the Boltzmann state~\eqref{boltzmann} are then exact to
  round-off. The matrix exponential of $\LL$ is formed from its symmetrised
  eigendecomposition.
\item \emph{Axial step.} For frozen $\eta$, \eqref{eq:chat} is solved exactly
  by the Ornstein--Uhlenbeck propagator,
  \begin{equation}
    \hat c(Y,t+\Delta t)=\hat c(Y+v\Delta t,\,t)\,
    \exp\!\Big[-\delta\,e^{2Y}\,\frac{e^{2v\Delta t}-1}{2v}\Big],
    \label{eq:OU}
  \end{equation}
  with four-point (cubic) Lagrange interpolation for the shift. The
  transverse step is applied over a full $\Delta t$, flanked by two axial
  half-steps (Strang splitting).
\item \emph{Inversion.} The concentration is recovered from
  $c=\pi^{-1}\!\int_0^\infty\hat c\cos(kz)\,dk$ in the tube and
  $c=\int_0^\infty k\,J_0(k\xi)\,\hat c\,dk$ for the surface flow. With
  $s=k\xi$ and a Gaussian window $W(s)=e^{-(\epsilon s)^2/2}$, $\epsilon=0.05$,
  the tube result is
  \begin{align}
    &c(\eta,\xi,t)=\frac{1}{\pi\xi}\int \hat c\big(\eta,\ln s-\ln\xi,t\big)\,
    s\cos s\,W(s)\,\mathrm{d}\ln s,
   \\
    &\frac{\mathrm{d}M}{\mathrm{d}X}=\frac{2}{\pi}\int \bar{\hat c}\,
    s\cos s\,W(s)\,\mathrm{d}\ln s,
    \label{eq:inversion}
  \end{align}
  where $\bar{\hat c}=\int_{\Omega_\eta}\hat c\,\mathrm{d}\eta$. The factor $1/\pi$ is that of the cosine inversion, and $S_1=2$ counts $\pm\xi$; no area factor $\pi$ enters because $M_{\rm tot}$ is defined without it. The window
  amounts to smoothing $c$ over a relative length $\epsilon$ in $\xi$, which is
  far below the scale on which $c$ varies in $\ln|\xi|$. It removes the
  oscillatory cancellation, except far into the forward tail, where values
  below $10^{-5}$ of the peak are discarded. In the plateau, $c$ at
  $|\xi|\ll e^{x_*t}$ is built from $\hat c$ at $k\sim e^{-x_*t}$, so the
  quadrature in $\ln s$ must extend down to $Y_{\min}$; this requires a
  coarser grid for $s\ll1$, where $\cos s\simeq1$.
\end{enumerate}
The grid in $Y$ has spacing $0.05$ and spans
$[-(v_{\max}t_{\max}+10),\ \ln(12/\Sigma_0)]$. We use $128$ and $160$
transverse cells and $\Delta t=0.02$ and $0.004$ for $\Pec=1$ and $10$,
respectively.

\subsection{Asymptotic predictions without fitting}

The physical quantities $\Lambda(q)$, $\phi_q$, $\Lambda'$ and $\Lambda''$ are computed from the
principal eigenpair of the same discrete operator $\LL+qv$ (with $\Lambda'$ from
the Hellmann--Feynman formula~\eqref{eq:variational} and $\Lambda''$ from second-order
perturbation theory), so that spatial discretisation error cancels in the
comparison. 

Unlike the exponents, widths and cross-sections, the amplitude $a(q)$
in~\eqref{eq:mellin_eq} is not part of the asymptotic theory, but for the present
initial condition it can be obtained exactly. Neglecting longitudinal diffusion,
$\mathcal M_q$ obeys~\eqref{eq:moments} with the initial datum
$g(q)\cdot 1$, where
\begin{equation}
  g(q)=\int_0^\infty \frac{e^{-\xi^2/2\Sigma_0^2}}{\sqrt{2\pi}\,\Sigma_0}\,\xi^{q}\,\mathrm{d}\xi
      =\frac{2^{q/2}\Sigma_0^{q}\,\Gamma\!\big(\tfrac{q+1}{2}\big)}{2\sqrt{\pi}}
\end{equation}
is the one-sided Mellin transform of the release (it has a simple pole at $q=-1$
with residue $1/\sqrt{2\pi}\Sigma_0$). Projecting the uniform initial
cross-section onto the principal mode, using that $\LL+qv$ is self-adjoint in
the $m_\infty$-weighted inner product with eigenfunction $h_q$
($\phi_q=m_\infty h_q/\!\int m_\infty h_q\,\mathrm{d}\eta$), gives
\begin{equation}
  a(q)=g(q)\,P(q),\qquad
  P(q)=\frac{\int_{\Omega_\eta} h_q\,\mathrm{d}\eta\ \int_{\Omega_\eta} m_\infty h_q\,\mathrm{d}\eta}
            {\int_{\Omega_\eta} m_\infty h_q^2\,\mathrm{d}\eta},
\end{equation}
so that $a(0)=\tfrac12=M_{\rm tot}/S_d$ and
$K=P(-d)/(\sqrt{2\pi}\,\Sigma_0)$. The predictions plotted below are the
plateau and erfc layer, \eqref{eq:plateau} and \eqref{eq:transition}, for
$x\le x_*+O(t^{-1/2})$, and the saddle-point results \eqref{eq:tail} and
\eqref{eq:massdensity} beyond; no parameter is adjusted. The only approximation in $a(q)$ is the neglect of the early longitudinal
diffusion, which changes the plateau amplitude by a relative amount of order
$\delta/U\Sigma_0^2$. All the quantities $U$, $D$, $\Lambda(-d)$ and $x_*$
obtained from the discrete operator agree with Table~\ref{tab:Lambda} to within
$10^{-3}$.

\begin{figure}[h!]
  \centering
  \includegraphics[scale=0.5]{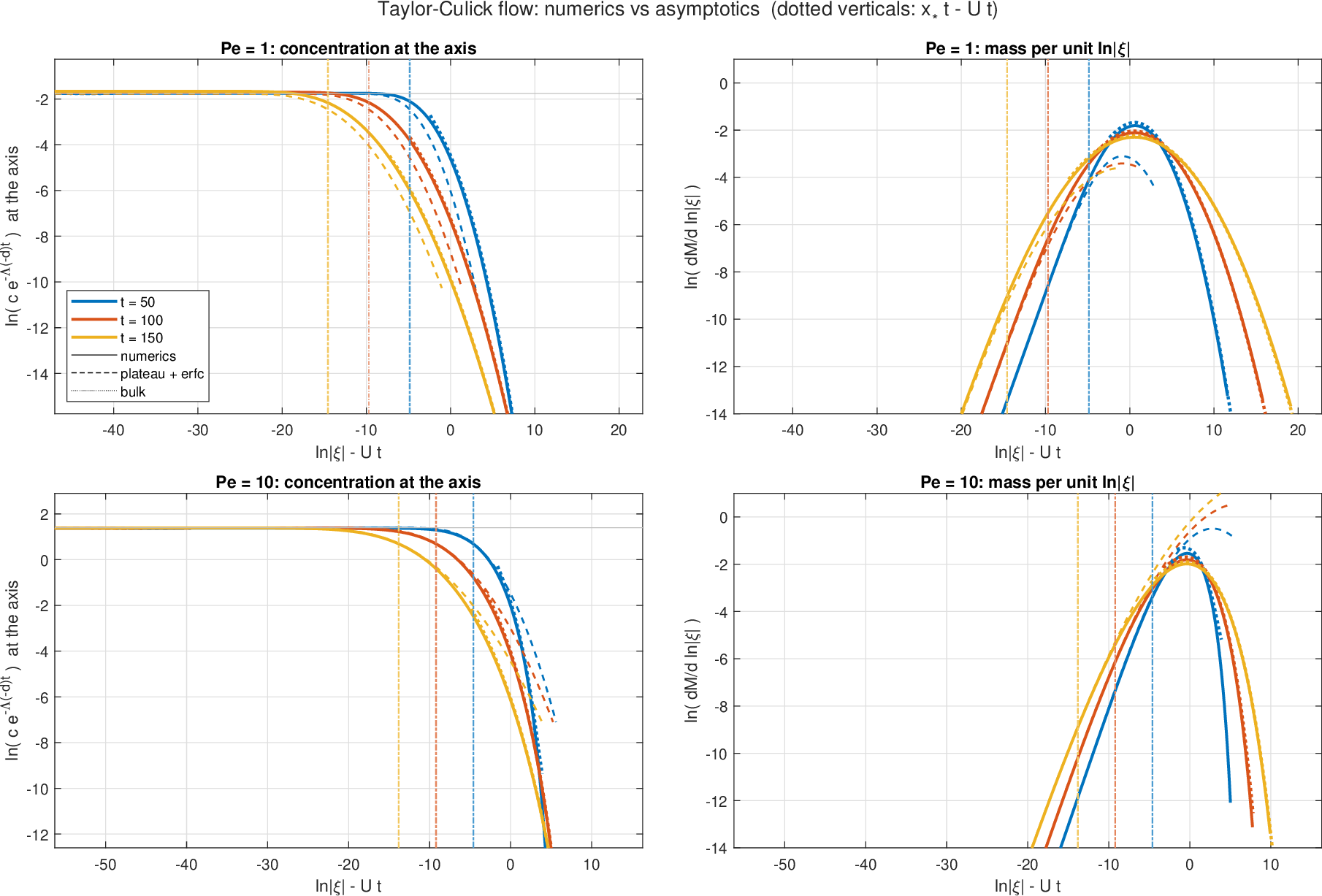}
  \caption{Numerical solution versus asymptotic structure for the Taylor--Culick tube flow at $\Pec=1$ (top; $t=50,100,150$) and
  $\Pec=10$ (bottom; $t=5,10,15$). The initial release is a Gaussian in $z$ of width $\Sigma_0=3$ ($\Pec=1$) or $1$ ($\Pec =10$), uniform in $\eta$, with $M_{\rm tot}=1$. Left:
  concentration at the cell nearest the axis, $\ln(c\,e^{-\Lambda(-d)t})$.
  Right: mass per unit logarithmic length, $\ln(\mathrm{d}M/\mathrm{d}\ln|\xi|)$.
  Both are plotted against $\ln|\xi|-Ut$. Solid: numerical solution; dashed:
  plateau and erfc layer, \eqref{eq:plateau} and \eqref{eq:transition}; dotted:
  saddle-point results, \eqref{eq:tail} and \eqref{eq:massdensity}, with $a(q)$
  computed from the initial condition. Dash-dotted verticals mark the plateau edge
  $x_*t-Ut$. The horizontal grey line is the plateau level
  $\ln[K\phi_{-d}(0)]$.}
  \label{fig:culick}
\end{figure}

\subsection{Results}

Figure~\ref{fig:culick} compares the numerical solution with the asymptotic
structure of figure~\ref{fig:sch} at three times for each Peclet number
($t=50,100,150$ for $\Pec=1$ and $t=5,10,15$ for $\Pec=10$, chosen so that the
separation $(U-x_*)t$ between plateau and bulk is similar, about $5$--$15$ in
$\ln|\xi|$). The left panels show the concentration at the cell nearest the
axis, scaled by the plateau decay, $\ln(c\,e^{-\Lambda(-d)t})$, and the right
panels show the mass per unit $\ln|\xi|$; both are plotted against
$\ln|\xi|-Ut$, so that the bulk is centred at the origin and the dash-dotted
verticals mark the edge of the plateau, $(x_*-U)t$.

\emph{Plateau and transition zone.} After scaling by $e^{\Lambda(-d)t}$ the
axis concentration collapses onto a single horizontal level, equal to
$\ln[K\phi_{-d}(0)]$, for all three times. This confirms both the decay rate
$-\Lambda(-d)$, which differs substantially from $-dU$ (Table~\ref{tab:Lambda}), and
the independence of $c$ on $\xi$. The plateau extends over a range of
$\ln|\xi|$ that grows linearly in time. It is bounded on the right by an
erfc layer centred at $x_*t$, whose width grows only like $\sqrt{t}$, as seen
from the dashed curves. The small displacement that remains between the
numerical layer and the dashed erfc curve is of order unity in $\ln|\xi|$; it
is expected, since~\eqref{eq:transition} neglects the cubic term of $\Lambda$ near
$q=-d$ and the regular part $\breve a$, both of which are of relative order
$t^{-1/2}$ compared with the layer width.

\emph{Rear tail, bulk and forward tail.} To the right of the transition zone,
the saddle-point results (dotted) lie on the numerical curves, through the rear
tail, the bulk and the forward tail, in both panels. In the concentration panels
the slope of $\ln c$ against $\ln|\xi|$ steepens continuously from $-d$ at the
transition zone, as predicted by $-(q_s+d)$. The mass panels show the same
quantity as a log-normal-like profile whose  peak lies at $\ln|\xi|-Ut=X_0=\mc O(1)$. There the
saddle point is $q_s=O(t^{-1/2})$, so the cross-section is $m_\infty(\eta)$. The profile is visibly non-Gaussian away from
the peak: the two tails have different curvature, which the quadratic
closure $\Lambda\simeq Uq+Dq^2$ cannot represent, and which is captured by the
full $\Lambda(q)$. This is most evident for $\Pec=10$, where
$\varPi=1.08>1$ (Table~\ref{tab:Lambda}).

The Boltzmann state is preserved to $5\times10^{-14}$ and the mass
$\hat c(k\to0)$ to six digits at all times. The smoothing window changes the
integrated mass by $O(\epsilon^2)$; we find $\int(\mathrm{d}M/\mathrm{d}X)\,\mathrm{d}X=1.0025=1+\epsilon^2$ in all cases. As a
stringent test of the plateau, we compare the numerical amplitude
$K_{\rm num}=c(\eta{=}0,\xi{=}0,t)\,e^{-\Lambda(-d)t}/\phi_{-d}(0)$ with the
predicted $K$. For $\Pec=10$ the ratio is $0.980$, $0.983$, $0.986$ at
$t=5,10,15$, and for $\Pec=1$ it is $1.016$, $1.060$, $1.107$ at
$t=50,100,150$. The deviation from unity at the earliest times is
consistent with the neglected early diffusion, of relative order
$\delta/U\Sigma_0^2$. The slow drift with time is an exponential effect:
$\ln(K_{\rm num}/K)$ grows linearly in $t$ at the rate
$8.6\times10^{-4}$ ($\Pec=1$) and $6.6\times10^{-4}$ ($\Pec=10$), i.e. the
numerical plateau decays at a rate that differs from $-\Lambda(-d)$ by only
$0.04\%$ and $0.002\%$, respectively. We attribute this to the accumulated
time-discretisation error of the splitting and of the interpolation,
rather than to the asymptotics, by computations carried out for different values of the time step.

\section{Concluding remarks}\label{sec:conclu}

In a linearly accelerating flow whose strain rate $v(\eta)$ varies across the section,
exponential stretching outpaces transverse diffusion, so the scalar never homogenises and a
cross-sectionally averaged description is not closed. In the coordinate $X=\ln|\gb\xi|$,
however, the transport equation~\eqref{eq:log_transport} is a Taylor-dispersion problem without
molecular diffusion along $X$: the strain rate $v(\eta)$ plays the role of the axial velocity,
and the cross-section relaxes to the Boltzmann profile $m_\infty(\eta)$. The mean
log-stretching rate $U$ and the log-space diffusivity $D$ are given by the
quadratures~\eqref{eq:U} and~\eqref{eq:D}, and $D>0$ only if $v$ varies across the section. Taylor
dispersion therefore emerges without homogenisation.

This is only the centre of the cloud. The principal eigenvalue $\Lambda(q)$ of $\LL+qv$
determines the whole large-time structure (figure~\ref{fig:sch}): a log-normal bulk of variance
$2Dt$ in $\ln|\gb\xi|$ that carries essentially all the scalar mass; forward and rear tails in
which the mass per unit $\ln|\gb\xi|$ is $e^{-I(x)t}$ to leading exponential order (relative
error $O(t^{-1})$ at fixed $x$); and, behind the rear tail, a plateau independent of $\gb\xi$
that decays as $e^{\Lambda(-d)t}$ with cross-section $\phi_{-d}(\eta)$, joined to the tail by a
complementary-error-function layer of width $O(\sqrt t)$. The plateau has the highest
concentration but only a fraction $e^{-I(x_*)t}$ of the mass. A probe near the stagnation point
therefore records the plateau, whose decay rate $-\Lambda(-d)$ is set by the whole cross-section
and is neither the maximum strain rate nor $dU$, whereas the bulk, at a distance $(U-x_*)t$ in
$\ln|\gb\xi|$, shows where the scalar is. The amplitudes $K$ and $a(q)$ and the shift $X_0$
carry the memory of the release and are the only parts of the solution that depend on it.

The log-normal bulk and the tails governed by $\Lambda(q)$ at $q\neq0$ are reminiscent of the
statistics of material stretching in turbulent strain-dominated
regions~\cite{kraichnan1974convection,girimaji1990material,falkovich2001particles}. Here the
log-normal form arises from transverse exchange between regions of different strain, with a
finite log-space diffusivity $D$, and the departures from it are given by the full
$\Lambda(q)$. Whether a plateau and an erfc front have counterparts in turbulence depends on the
existence of a fixed stagnation point and on the tensorial and rotational character of the
strain, which the scalar model here does not capture. The analysis assumes a steady strain rate $v(\eta)\ge0$ that is linear in $\gb\xi$ with a
coefficient depending only on $\eta$, and several extensions lie outside it.
\cite{brady1981steady} found solutions of the same similarity form for $\Rey<0$ in which the
strain rate becomes negative over part of the cross-section. Where it is negative the scalar is compressed along $\gb\xi$ rather
than stretched, longitudinal diffusion is no longer negligible (Section~\ref{sec:justify}), and the
sign argument behind $\Lambda'>0$ fails, so the structure found here would have to be reworked.
Whether such flows remain stable at large $|\Rey|$ is a separate question that we do not address.
Unsteady strain~\cite{krishnan1992diffusion} would make the generating function~\eqref{eq:moments}
time dependent, and other stretched flows, with a different geometry or without a self-similar
form, may require a different logarithmic or algebraic coordinate. We leave these cases to future work.

\vspace{1cm}

\appendix

{\centering\Large Appendix\par}

\section{The tube problem at large Peclet numbers}\label{app:largepe}
For $\Pec\gg1$ the Boltzmann profile~\eqref{boltzmann} is $m_\infty\simeq r\,e^{-r\eta}$, with
$r\equiv\Pec f_0'$ and $f_0'=f'(0)$, so the scalar is confined to a layer $\eta=O(\Pec^{-1})$
about the axis. In this layer
\begin{equation}
  f=f_0'\eta+\tfrac12f_0''\eta^2+\tfrac16f_0'''\eta^3+\cdots,\qquad
  v=4\Pec f'=4\Pec f_0'-\chi\,\eta+O(\Pec\,\eta^2),\qquad \chi\equiv4\Pec|f_0''| ,
\end{equation}
where $f_0''\le0$. For the Taylor--Culick flow $f_0'=\pi/4$, $f_0''=0$ and $f_0'''=-\pi^3/16$.

\emph{Transverse relaxation.} With $\zeta=\Pec\,\eta$ and $f\simeq f_0'\eta$, the adjoint
problem~\eqref{eq:relax} becomes $\zeta g_{\zeta\zeta}+(1-f_0'\zeta)g_\zeta+(\lambda/4\Pec)\,g=0$,
with $g'(\infty)\to0$ (polynomial growth); equivalently
$\LL^\dagger g=4\eta g''+4(1-r\eta)g'=-\lambda g$. With $s=f_0'\zeta=r\eta$, this is Laguerre's
equation $sg_{ss}+(1-s)g_s+n\,g=0$, with $n=\lambda/(4\Pec f_0')$, whose polynomial solutions
require $n=0,1,2,\dots$:
\begin{equation}
    g_n=L_n(r\eta),\qquad \lambda_n=4n\,\Pec f_0' .
    \label{eq:laguerre}
\end{equation}
The first relaxation rate is $\lambda_1=4\Pec f_0'$, i.e.\ $\pi\Pec$ for the Taylor--Culick flow
(Table~\ref{tab:Lambda}: $\lambda_1=29.0$ against $\pi\Pec=31.4$ at $\Pec=10$).

\emph{Plateau problem ($\LL-v$ acting on $c\propto e^{-\mu t}F$).} The equation is
$(\eta F')'+\Pec fF'+(\mu/4)F=0$. At leading order, $\Pec fF'+(\mu/4) F=0$, so
$F=C\exp\big(\frac{\mu}{4\Pec}\int_\eta^1 d\hat\eta/f(\hat\eta)\big)$; this cannot satisfy the wall
condition~\eqref{eq:robin} (it would need $\mu=\Pec^2$), so $C=0$, hence $F=0$ in the outer region
and a thin layer remains at $\eta\sim1/\Pec$. In terms of $\zeta=\Pec\eta$ and $\varpi=\mu/(4\Pec)$,
\begin{equation}
    \frac{d}{d\zeta}\Big(\zeta\frac{dF}{d\zeta}\Big)+f_0'\zeta\frac{dF}{d\zeta}+\varpi F=0,
    \qquad \lim_{\zeta\to0}\sqrt\zeta F_\zeta=0,\quad F(\infty)\to0\ \text{(exponentially, to match the outer solution)}.
\end{equation}
The solution regular at the origin is $F\sim e^{-f_0'\zeta}M(1-\varpi/f_0',1,f_0'\zeta)$, with $M$
the confluent hypergeometric function. As $\zeta\to\infty$ it has an algebraic part
$\propto(f_0'\zeta)^{-\varpi/f_0'}/\Gamma(1-\varpi/f_0')$ and an exponential part
$\propto e^{-f_0'\zeta}(f_0'\zeta)^{\varpi/f_0'-1}/\Gamma(\varpi/f_0')$; requiring exponential decay
sets $1/\Gamma(1-\varpi/f_0')=0$, i.e.\ $\varpi_n=f_0'(1+n)$ and $F_n=e^{-f_0'\zeta}L_n(f_0'\zeta)$.
The leading mode $\varpi_0=f_0'$, $F_0=e^{-f_0'\zeta}$ corresponds to $\mu_0=4\Pec f_0'$, i.e.\
$\Lambda(-1)=-4\Pec f_0'+O(1)$. This eigenvalue selection is similar
to that done in~\cite{robinson1984stability,rajamanickam2021steady}.

\emph{Closed form for $\Lambda(q)$.} Take the model problem $f=f_0'\eta$ in the drift and
$v=4\Pec f_0'-\chi\eta$, on $0\le\eta<\infty$. Substituting $c=e^{-b\eta}$ into
$(\LL+qv)c=\Lambda c$, the coefficients of $\eta$ give $4b^2-4\Pec f_0'b-q\chi=0$, i.e.\
$b=\tfrac12\big[\Pec f_0'+\sqrt{\Pec^2f_0'^2+q\chi}\big]$ (the root with $b\to\Pec f_0'$ as $q\to0$,
the Boltzmann profile), and the constant terms give $\Lambda=4\Pec f_0'q+4\Pec f_0'-4b$, i.e.
\begin{equation}
    \Lambda(q)=4\Pec f_0'\,q+2\Pec f_0'\Big[1-\sqrt{1+\frac{4q|f_0''|}{\Pec f_0'^2}}\Big],
    \qquad U\to4\Pec f_0'-\frac{4|f_0''|}{f_0'} .
    \label{eq:Lambdalargepe}
\end{equation}
This is exact for the model problem. For the tube it holds to $\mc O(1)$, with absolute error
$\mc O(\Pec^{-1})$ at fixed $q$, because the neglected terms ($\mc O(\Pec\,\eta^2)$ in $v$, $\mc O(f_0'')$ in
the drift, and the wall condition, which acts only at exponentially small level) contribute at
$\mc O(\Pec^{-1})$; the square root matters only if $4|q||f_0''|/(\Pec f_0'^2)=\mc O(1)$. At $q=-1$ it
requires $\Pec f_0'^2>4|f_0''|$; below this the eigenfunction is no longer confined to the layer
and spreads across the tube.

\emph{Taylor--Culick flow.} Here $\chi=0$, so~\eqref{eq:Lambdalargepe} gives $\Lambda=4\Pec f_0'q$,
and $U$ and $D$ come from the next order in the layer expansion. With $x=r\eta$ and
$r=\pi\Pec/4$, $m_\infty\,\mathrm d\eta=e^{-x}\mathrm dx$ and
\begin{equation}
    v=\pi\Pec\cos\tfrac{\pi\eta}{2}=\pi\Pec-\frac{2\pi}{\Pec}\,x^2+\mc O(\Pec^{-3}),\qquad
    x^2=2L_0-4L_1+2L_2 ,
\end{equation}
so that $v-U=\sum_{n\ge1}c_nL_n(x)$ with $c_1=8\pi/\Pec$, $c_2=-4\pi/\Pec$, and
$U=\pi\Pec-4\pi/\Pec$. The Laguerre polynomials are orthonormal with weight $e^{-x}$ and are the
eigenfunctions~\eqref{eq:laguerre} of $\LL^\dagger$. Since $D=\langle v\,h_1\rangle$ with
$\LL^\dagger h_1=U-v$ (Appendix~\ref{app:taylor}), $D=\sum_{n\ge1}c_n^2/\lambda_n$, so that
\begin{equation}
    U=\pi\Pec-\frac{4\pi}{\Pec}+\mc O(\Pec^{-3}),\qquad
    D=\frac{64\pi}{\Pec^{3}}+\frac{8\pi}{\Pec^{3}}=\frac{72\pi}{\Pec^{3}}\big[1+\mc O(\Pec^{-2})\big].
    \label{eq:UDlargepe}
\end{equation}
The strain varies across the occupied layer by only $\mc O(\Pec^{-1})$, while the relaxation rate is
$\mc O(\Pec)$, which is why $D\sim\Pec^{-2}/\Pec$. Each further cumulant brings a factor
$\mc O(q/\Pec^{2})$, so that $\Lambda(q)=Uq+Dq^2+\mc O(q^3\Pec^{-5})$, and the quadratic closure becomes
exact at large $\Pec$.

\section{Taylor-expansion coefficients of $\Lambda(q)$ to all orders} \label{app:taylor}

The coefficients of $\Lambda(q)=\sum_{n\ge1}\Lambda_nq^n$ follow from a recursion of quadratures.
They are the cumulants of the accumulated log-stretch, $\Lambda_n=\kappa_n/n!$, where the $n$th
cumulant of $\ln|\gb\xi|$ grows as $\kappa_nt$ (so $\kappa_1=U$ and $\kappa_2=2D$), and $\Lambda_3$
gives the skewness correction to the log-normal bulk (Section~\ref{sec:gauss}). Write
$h_q=1+\sum_{n\ge1}q^nh_n$ with $\langle h_n\rangle\equiv\int_{\Omega_\eta}m_\infty h_n\,\mathrm d\eta=0$
for $n\ge1$. At $\mc O(q^n)$, \eqref{eq:eig_h} gives
\begin{equation}
  \left(\alpha\,m_\infty h_n'\right)'=m_\infty\Big[\sum_{k=1}^{n}\Lambda_k h_{n-k}-v\,h_{n-1}\Big],\qquad h_0=1 .
  \label{eq:recursion}
\end{equation}
Integrating~\eqref{eq:recursion} over $\Omega_\eta$, the left side vanishes by the flux condition
$h_n'=0$, and since $\langle h_j\rangle=0$ for $j\ge1$ one obtains $\Lambda_n=\langle v\,h_{n-1}\rangle$.
One further quadrature gives
\begin{equation}
  h_n'(\eta)=\frac{1}{\alpha\,m_\infty}\int_0^\eta m_\infty\Big[\sum_{k=1}^{n}\Lambda_k h_{n-k}-v\,h_{n-1}\Big]\mathrm{d}s ,
\end{equation}
with the constant in $h_n$ fixed by $\langle h_n\rangle=0$. For $n=1$ this gives
$h_1'=-W/(\alpha m_\infty)$ and $\Lambda_1=U$, and for $n=2$,
$\Lambda_2=\langle v h_1\rangle=\int_{\Omega_\eta}W^2/(\alpha m_\infty)\,\mathrm d\eta=D$, which are
the expressions \eqref{eq:U} and \eqref{eq:D}. The first correction to the eigenfunction is
$\psi_1=m_\infty h_1$, that is $\psi_1'+\beta\psi_1=-W/\alpha$. Each $\Lambda_n$ is thus an iterated
quadrature of the strain profile and requires no further eigenvalue problem.

\emph{Size of the coefficients.} Each step of the recursion applies $(-\LL^\dagger)^{-1}$, which
divides by a relaxation rate of order $\lambda_1$, to a factor $v-U$ of the order of the strain
contrast $v_{\max}-v_{\min}$. Hence $h_n=O(\Pi^n)$ and $\Lambda_n=O(\lambda_1\Pi^n)$ for $n\ge2$,
up to numerical prefactors, with $\Pi=(v_{\max}-v_{\min})/\lambda_1$ (Section~\ref{sec:gauss}). The series
in $q$ is therefore controlled by $|q|\,\Pi$, and the quadratic closure $\Lambda\simeq Uq+Dq^2$ is
accurate at $q=-d$ when $d\,\Pi\ll1$. This is a scaling estimate, not a bound. The prefactors are
small ($D\simeq0.07\,\Pi^2\lambda_1$ for the Taylor--Culick tube at $\Pec=1$ and
$0.06\,\Pi^2\lambda_1$ for the Homann flow at $\Sch=1$), so it is a conservative criterion.
It also overstates the effect when the cloud samples only a part of the strain range: for the
Taylor--Culick tube at large $\Pec$, $\Pi\to1$ while the closure becomes exact
(Appendix~\ref{app:largepe}). Where $d\,\Pi\gtrsim1$ the series is not needed, because
$\Lambda(-d)$ is computed directly from~\eqref{eq:eig}.

% --- Bibliography Section ---
% References will show up cleanly under a standard numbered layout
\bibliographystyle{unsrt}
\bibliography{references} % Points to your references.bib file

\end{document}